\documentclass[aps,preprint,nofootinbib,preprintnumbers,eqsecnum,superscriptaddress]{revtex4}
\usepackage{color}

\newcommand{\Comment}[1]{{}}
\usepackage[
      colorlinks=true,
      linkcolor=blue,
      urlcolor=blue,
      filecolor=black,
      citecolor=red,
      pdfstartview=FitV,
      pdftitle={},
        pdfauthor={Arpit Das, Nabil Iqbal, Napat Poovuttikul},
        pdfsubject={},
        pdfkeywords={},
        pdfpagemode=None,
        bookmarksopen=true
      ]{hyperref}

\usepackage[font=footnotesize,labelfont=bf,justification=centerlast,width=.94\textwidth]{caption}

\usepackage{threeparttable}
\usepackage{tikz}

\usepackage[normalem]{ulem}
\usepackage{amsmath}
\usepackage{enumerate}
\usepackage{amsfonts}
\usepackage{yfonts}
\usepackage{array,xcolor,graphicx}

\usepackage{subfigure}
\usepackage{psfrag}
\usepackage[compat=1.0.0]{tikz-feynman}
\usepackage{epsfig}
\usepackage[latin1]{inputenc}
\usepackage{float}
\usepackage{dsfont}
\usepackage{graphicx}
\usepackage{caption}
\usepackage{subcaption}
\usepackage{cancel}
\usepackage{mathrsfs}
\usepackage{amssymb}
\usepackage{amsfonts}
\usepackage{amsmath}
\usepackage{framed}
\usepackage{slashed}
\usetikzlibrary{positioning}
\usepackage{amsmath}
\usetikzlibrary{arrows}
\usepackage{pgfplots}
\usetikzlibrary{calc,fadings,decorations.pathreplacing}
\usepackage{tikz-3dplot}
\tdplotsetmaincoords{105}{0}
\pgfmathsetmacro{\ax}{1}
\pgfmathsetmacro{\ay}{0}
\pgfmathsetmacro{\az}{1}

\newcommand \tr {\mbox{{\bf Tr}}}

\usepackage{graphicx}
\usepackage{bm}

\def\({\left(}
\def\){\right)}
\def\[{\left[}
\def\]{\right]}
\def\<{\langle}
\def\>{\rangle}
\newcommand{\bmat}{\begin{bmatrix}}
\newcommand{\emat}{\end{bmatrix}}

\def\Tr{\mathop{\rm Tr}}
\def\tr{\mathop{\rm tr}}

\newcommand\half{{\ensuremath{\frac{1}{2}}}}
\newcommand\p{\ensuremath{\partial}}

\newcommand\vev[1]{{\ensuremath{\left\langle{#1}\right\rangle}}}

\newcommand\ket[1]{\ensuremath{\lvert{#1}\rangle}}
\newcommand\bra[1]{\ensuremath{\langle{#1}\rvert}}

\newcommand{\be}{\begin{equation}}
\newcommand{\ee}{\end{equation}}
\newcommand{\bea}{\begin{eqnarray}}
\newcommand{\eea}{\end{eqnarray}}
\newcommand{\bwt}{\begin{widetext}}
\newcommand{\ewt}{\end{widetext}}

\newcommand{\bi}{\begin{itemize}}
\newcommand{\ei}{\end{itemize}}
\newcommand{\ben}{\begin{enumerate}}
\newcommand{\een}{\end{enumerate}}
\newcommand{\bca}{\begin{cases}}
\newcommand{\eca}{\end{cases}}
\newcommand{\bln}{\begin{align}}
\newcommand{\eln}{\end{align}}
\newcommand{\bst}{\begin{split}}
\newcommand{\est}{\end{split}}

\newcommand\al{{\alpha}}
\newcommand\ep{\epsilon}

\newcommand\lam{\lambda}
\newcommand\Lam{\Lambda}
\newcommand\om{\omega}
\newcommand\Om{\Omega}

\newcommand\ga{{\ensuremath{{\gamma}}}}

\def\th{{\theta}}

\newcommand\ha{{\half}}

\def\le{\left}
\def\ri{\right}

\newcommand\sF{{\ensuremath{{\mathcal F}}}}

\newcommand\sO{{\ensuremath{{\mathcal O}}}}

\begin{document}

\title {Wilson lines with endpoints in 3d CFT}

\author{Nabil Iqbal}
    \email{nabil.iqbal@durham.ac.uk}
    \affiliation{Centre for Particle Theory, Department of Mathematical Sciences, Durham University,
		South Road, Durham DH1 3LE, UK\\} 
    \affiliation{Amsterdam Machine Learning Lab, University of Amsterdam,
Science Park 900, 1098 XH Amsterdam, NL}
\author{Navonil Neogi}
	\email{navonil.neogi@durham.ac.uk}
	\affiliation{Centre for Particle Theory, Department of Mathematical Sciences, Durham University,
		South Road, Durham DH1 3LE, UK\\} 
        \affiliation{Institute for Theoretical Physics, University of Amsterdam, Science Park 904, 
    1098 XH Amsterdam, NL}
   
\begin{abstract}
 In Abelian gauge theories with dynamical matter, Wilson lines can end on the insertions of charged fields. We study the endpoints of Wilson lines in large $N$ bosonic QED$_3$ at its critical point. We first study the stability of an infinite Wilson line in the $\mathbb{CP}^{N-1}$ model by computing the appropriate functional determinant at large $N$. We also compute the conformal dimension of the lowest-dimension endpoint of the line to first order in $N^{-1}$. Along the way we calculate the field-strength tensor $F_{\mu\nu}$ in the presence of the line with endpoint and discuss a state-operator correspondence for the endpoints, as well as the existence of an OPE that allows two open-ended Wilson lines to be glued together into a single line. 

\end{abstract}

\vfill

\today
    \hypersetup{linkcolor=black}

\maketitle

\tableofcontents
\vspace{2cm}

\section{Introduction}

What is an electron?

In elementary quantum field theory we often say that an electron is the object that is created by a quantum field which we write $\psi(x)$. This is of course not the entire story; the electron is charged under a gauge symmetry, which means that the physical electron cannot be considered an independent entity from the electric field that it creates. Mathematically this means that we must work with the gauge-invariant object $\psi(x)e^{i\int_{x} A}$, where one can imagine that $\psi$ creates the electron and the associated Wilson line $e^{i\int_{x} A}$ creates its electric field. From this point of view one might even be tempted to view the electron as being {\it defined} as the object which terminates an electric field line. 

This is not a useful point of view\footnote{At least we don't think so.} for real-life electrons, but it resonates nicely with recent ideas associated with higher-form symmetry in quantum field theory, where gauge theories without charged matter are understood as having a {\it 1-form symmetry} \cite{Gaiotto:2014kfa}\footnote{See e.g. \cite{Cordova:2022ruw,Brennan:2023mmt,Gomes:2023ahz,Shao:2023gho,Schafer-Nameki:2023jdn,Bhardwaj:2023kri,McGreevy:2022oyu,Iqbal:2024pee} for recent reviews on this subject.}. which ensures the existence of a conserved electric flux. The addition of charged matter means that the 1-form symmetry is explicitly broken. Thus the charged matter itself should be characterized by the fact that it is the deformation that breaks the symmetry, i.e. by the fact that it terminates the electric field. 

In this work we attempt to think about charged matter in this fashion, i.e. by considering it as the endpoint of a Wilson line. For this to be an interesting point of view the dynamics of the gauge field should be strong enough -- or difficult enough to disentangle from the charged degrees of freedom -- that viewing the two as a composite object makes sense. We thus study this in the simplest dynamically non-trivial setting where this is the case, the $\mathbb{CP}^{N-1}$ model, i.e. the CFT fixed point of QED$_{3}$ coupled to $N$ bosonic flavors. In this case the IR dynamics of the gauge field arises entirely from its interaction with the charge-carriers. Furthermore, the Wilson line itself flows to a conformal defect: see e.g. \cite{Billo:2016cpy} for a review and \cite{Billo:2013jda,Bianchi:2022sbz,Cuomo:2021kfm,Gaiotto:2013nva,Pannell:2023pwz} for  work describing line defects in conformal field theory. Wilson lines in U(1) theories were studied in \cite{Metlitski:2007fu}. Of particular relevance to our study is recent work describing the rich dynamics of Abelian Wilson lines from this point of view \cite{Aharony:2022ntz,Aharony:2023amq}. We will build on this by studying the properties of the {\it endpoint} of the line from the point of view of defect conformal field theory. 

In more detail we will study Abelian Wilson lines with endpoint of the following form:
\be
W(C;x) = \exp\le(i\int_{x} A\ri) \phi_{\al}(x)
\ee
where the curve $C$ ends at the point where the scalar field $\phi$ is inserted at $x$. One of our main results will be to compute the conformal dimension of the endpoint of the line explicitly to first order in the large $N$ expansion to be $\Delta_{\phi} = \ha - \frac{18}{N \pi^2}$. 

We will take a leisurely route to get there; along the way we in Section \ref{sec:stateop} we review a state-operator correspondence for line-defect CFTs, where states on a punctured sphere are mapped to defect-localized operators, as previously described in \cite{Zhou:2023fqu} where it was used to study line defects in the 3d Ising CFT. Similar Wilson lines in Chern-Simons theories were studied in \cite{Gabai:2022mya, Gabai:2022vri}, in which the authors treat the endpoints of such lines as conformal objects and compute conformal dimensions and spectra. We pursue the same studies in bosonic $QED_3$ and the $\mathbb{CP}^{N-1}$ model in our work. 

We also explain the existence of an OPE expansion that allows one to sew together two open-ended Wilson line into a single closed Wilson line. The existence of similar relations was first discussed in the context of (bulk) applications to quantum gravity in \cite{Harlow:2015lma}. In Section \ref{sec:wilsonnoend} we discuss the stability of the infinite Wilson line, and in Section \ref{sec:wilsonend} we discuss various properties of the endpoint, including the profile of the field strength tensor near the endpoint of the line. In Section \ref{sec:confdim} we compute the Feynman diagrams contributing to the conformal dimension of the endpoint of the line. Most calculational details are relegated to the Appendices.

%\subsection{Background}

\subsection{QED in three dimensions}
We begin with a brief review of the theory that we study, bosonic scalar QED in three dimensions. This theory (henceforth referred to as QED$_3$) can flow to multiple IR fixed points and is very well-studied in a large $N$ expansion; see e.g. \cite{Coleman:1985rnk} for a textbook treatment and \cite{Benvenuti:2019ujm,kaul2008quantum,Benvenuti:2018cwd} for explicit discussion of the Feynman rules in various variants of the theory. %The fate of Wilson lines in these different limits was discussed in \cite{Aharony:2022ntz}, and we pick up from this discussion. 

 The action of bosonic QED with $N$ scalars, each with mass $m$ and unit charge, in 3 dimensions on $\mathbb{R}^3$ takes the following form:
\begin{equation}
S[\Phi, A] = \int_{\mathbb{R}^3} d^3 x \,\le( \sum_{\alpha=1}^N \left( |D_\mu \Phi_\alpha|^2  \right) +m^2 \sum_{\alpha=1}^N |\Phi_\alpha|^2 +\frac{1}{4e^2} F_{\mu\nu} F^{\mu\nu}\ri) 
\end{equation}
where the covariant derivative $D_\mu$ is defined as:
\begin{equation}
D_\mu \Phi_a := (\partial_\mu-iA_\mu) \Phi_a
\end{equation}
The quadratic Maxwell term in this action $F_{\mu\nu}^2$ is irrelevant in the IR limit of this theory which we shall consider later to arrive at the $\mathbb{CP}^{N-1}$ fixed point.

The 0-form global symmetries of this theory are $(PSU(N) \times U(1)_{m}) \rtimes \mathbb{Z}_2$, where the $PSU(N)$ flavor symmetry is generated by rotations of the $\phi \to U \phi$ of which the diagonal phase factor is part of the $U(1)$ gauge group and thus does not act on the physical space of states. The $U(1)_{m}$ symmetry is generated by the topological monopole current $j^{\mu} = \ep^{\mu\nu\rho} F_{\nu\rho}$ and the $\mathbb{Z}_2$ factor is charge conjugation; these will not play an important role in our analysis. 

If $m^2 > 0$ then in the IR the charged matter can be ignored and there is an emergent 1-form symmetry controlling conservation of electric flux. We will be interested precisely in the critical point where $m^2 = 0$, the charged matter is gapless, and this 1-form symmetry does not exist. 

To this gauge theory we may add $PSU(N)$ invariant interaction terms in the form of a potential $V(|\Phi_\alpha|^2)$. The theory may be tuned to a fixed point by appropriate tunings of the coupling, and is expected to flow to a CFT in the large $N$ limit. A review of these fixed points may be found in \cite{Benvenuti:2019ujm}. 

One such is the fine-tuned ``tricritical'' theory, which can be understood in the large-$N$ limit as fine-tuning both the mass $m=0$ and interaction terms $V(|\Phi_\alpha|^2) = 0$. There exists another fixed point where we add a quartic interaction of the form $V(|\Phi_\alpha|^2) = (g/2N) |\Phi_\alpha|^4$. %A standard procedure to deal with this new interaction is to introduce an auxiliary Hubbard-Stratonovich field $\lam$ to decouple the quartic interaction. The new action is given by:
%\begin{equation}
%S[\Phi, A, \lam] = \int_{\mathbb{R}^3} d^3 x \, \le(\sum_{\alpha=1}^N \left( |D_\mu \Phi_\alpha|^2  \right) +\lambda \sum_{\alpha=1}^N \Phi^\dag_\alpha \Phi_\alpha - \frac{N}{2g}\lambda^2 -\frac{1}{4e^2} F_{\mu\nu} F^{\mu\nu}\ri)
%\end{equation} 
This specific case is discussed in \cite{Benvenuti:2019ujm}\footnote{We find that it is helpful to consider the relationship between the tricritical and $\mathbb{CP}^{N-1}$ models as somewhat analogous to that between the free and Wilson-Fisher fixed points \cite{Klebanov:2002ja} for the non-gauged $O(N)$ vector model.}. At its critical point this theory is called the $\mathbb{CP}^{N_b-1}$ model, and can be written in terms of $N$ scalars that satisfy a length constraint, here enforced by a Lagrange multiplier $\lam$. The action on $\mathbb{R}_3$ takes the form
\be
S[\phi, A, \lam] = \int_{\mathbb{R}^3} d^3x \left( |D_\mu \Phi_\alpha|^2  -\frac{1}{4e^2} F_{\mu\nu} F^{\mu\nu} + \lambda \left(\sum_{\alpha=1}^N \left(|\Phi_\alpha|^2 \right) - \frac{N}{\frak{g}_c}\ri)\ri) \label{CPnR3} 
\ee
We will also often work in the conformal frame of $S^2 \times \mathbb{R}$, in which case the action takes the form:
\begin{equation}
\int_{S^2 \times \mathbb{R}} \sin \theta \, d\tau d\theta d\phi \, \le(\sum_{\alpha=1}^N \left( |D_\mu \Phi_\alpha|^2  +\frac{1}{4} |\Phi_\alpha|^2 \right) -\frac{1}{4e^2} F_{\mu\nu} F^{\mu\nu} + \lambda \left(\sum_{\alpha=1}^N \left(|\Phi_\alpha|^2 \right) - \frac{N}{\frak{g}_c} \right)\ri)
 \label{scalac} \end{equation}
where we note that the conformal coupling $\frac{1}{4} \Phi^2$ is specific to $S^2$.% we will switch between the $\mathbb{R}^3$ and $S^2 \times \mathbb{R}$ perspectives during the course of our calculations as convenience dictates.

\subsection{Wilson lines in QED$_3$}
A natural set of line operators in QED$_3$ are the Wilson lines of the Abelian gauge potential $A$. A Wilson line without endpoint is defined on a closed curve $C$ as
\be
W(C) = \exp\le(iq\int_{C} A\ri)
\ee
where the charge $q \in \mathbb{Z}$. If $C$ is chosen to be an infinite straight line or a circle then one expects these objects to flow to conformal defects at the CFT fixed point. These have been studied extensively and a rich dynamics of defect RG flows exists. 

It was argued in \cite{Aharony:2023amq} that these Wilson lines are unstable at the tricritical fixed point. At large $N$ the dimension of the operator $\sum_\alpha |\Phi_\alpha|^2$ is close to its free dimension of $1$. As this is a precisely marginal deformation; it is expected that any small electric field should further reduce its dimension, making it relevant and resulting in a runaway direction. We will explicitly see evidence for this instability below, and for this reason we will focus most of our attention on the $\mathbb{CP}^{N-1}$ model in what follows, where we demonstrate explicitly that it is cured.  

In gauge theories with no charged matter these line operators are charged under the an electric 1-form symmetry that measures electric flux, and thus the Wilson lines cannot end \cite{Gaiotto:2014kfa}. However, in QED$_3$ this 1-form symmetry is explicitly broken by the presence of charged matter which fluctuates strongly at the CFT fixed point. We thus may construct line operators with endpoints, which terminate on a scalar field insertion $\phi_{\al}(x)$. We write the unit-charge Wilson line with endpoint as:
\begin{equation}
W(x,\infty)_{\al} = \exp\le(iq\int_{x}^\infty A\ri) \phi_{\al}(x) \label{one-end-point} 
\end{equation}
It is clear that a gauge transformation $\phi(x) \rightarrow e^{i\alpha(x)} \phi(x)$, $A_\mu \rightarrow A_\mu + \partial_\mu \alpha$ leaves this object unchanged. The endpoint of this object transforms under a projective representation of $PSU(N)$. The generalisation to $q \neq 1$ can be found by taking appropriate products of $q$ $\phi_{\al}$'s and is discussed in detail in \cite{Aharony:2023amq}.

There is something slightly sleazy about the presentation \eqref{one-end-point}. Here we have presented a single endpoint at $x$, where the line stretched off to infinity, allowing us in principle to ignore a putative other endpoint. However, in a conformal theory the point at infinity is just a conventional point like any other: in particular, an inversion $x \to \frac{x}{|x|^2}$ brings it to a finite distance from all the other points. Thus one should worry about gauge-invariance at the other endpoint, and a more honest characterization of the simplest open-ended Wilson line is:
\be
W(x,\infty)_{\al\beta} = \phi_{\beta}^{\dagger}(\infty)\exp\le(iq\int_{x}^\infty A\ri) \phi_{\al}(x) \label{with-infinity}
\ee
This will be relevant in what follows. 
\section{Punctured sphere for line defects} \label{sec:stateop} 
We now present a brief review of the state-operator correspondence for line defects. For operators in the bulk of an ordinary CFT, it is very well understood that the spectrum of states in radial quantization (i.e. where we quantize on spatial slices of $S^2$) is in one-to-one correspondence with the set of local operators on $\mathbb{R}^3$. 

We now review how a similar state-operator correspondence works for operators that are localized on a conformal line defect \cite{Billo:2013jda,Billo:2016cpy}. Our notation will be specific to CFT$_3$ but can easily be adapted to any dimension if needed. Take coordinates $(x,y,z) = (x^{a},z)$ on $\mathbb{R}^{3}$ and consider a line defect along the line $z = 0$. If the defect is conformal then the system is invariant under the {\it defect conformal group} of $SO(2,1) \times SO(2)$, corresponding to conformal transformations along the line and $SO(2)$ rotations that mix $(x,y)$.

\begin{figure}
    \centering
    \begin{tikzpicture}
    
    % Define points
    \coordinate (A1) at (0,2);
    \coordinate (A2) at (0,0);
    \coordinate (B1) at (0,-2);
    \coordinate (B2) at (0,-4);
    
    % Draw lines
    \draw (A1) -- (A2);
    \draw (B1) -- (B2);
    \draw[dashed, thick] (A2) -- (B1);
    
    % Draw black dots at endpoints
    \filldraw[black] (A2) circle (2pt);
    \filldraw[black] (B1) circle (2pt);

    % Label the points
    \node[below left] at (A2) {$\phi^\dag(x)$};  % Label A
    \node[below left] at (B1) {$\phi(y)$};  % Label B

    \draw[thick] (0,-1) circle[radius=1.8cm];

    % Define points
    \coordinate (C1) at (4,2);
    \coordinate (C2) at (4,-4);
    \coordinate (C3) at (4,-1);
    
    % Draw lines
    \draw (C1) -- (C2);
    
    % Draw black dots at endpoints
    \filldraw[black] (C3) circle (2pt);

    \draw[thick] (4,-1) circle[radius=1.8cm];
    
    \end{tikzpicture}
    \begin{tikzpicture}
    % Set up the 3D plot with spherical coordinates
    \tdplotsetmaincoords{60}{120}  % view from an elevation of 60 degrees and azimuth of 110 degrees
    \begin{scope}[tdplot_main_coords]
        
        % Draw the sphere
        \shade[ball color=gray!50, opacity=0.5] (0,0,0) circle (3cm);
        
        % Draw the longitudinal circle (meridian)
        \tdplotdrawarc[thick]{(0,0,0)}{3}{10}{280}{}{}
        \tdplotdrawarc[dashed, thick]{(0,0,0)}{3}{10}{390}{}{}
        \filldraw[black] (2.954,0.521,0) circle (2pt);  % Top dot
        \filldraw[black] (0.521,-2.954,0) circle (2pt); % Bottom dot
        \filldraw[black] (-1.721,2.457,0) circle (2pt);
        \tdplotsetrotatedcoords{140}{90}{90}
        \tdplotgetpolarcoords{\ax}{\ay}{\az}
        %\tdplotsetthetaplanecoords{\tdplotresphi}
        \tdplotdrawarc[tdplot_rotated_coords, dotted, thick]{(0,0,0)}{3}{0}
        {360}{anchor=east}

    \end{scope}
    \end{tikzpicture}
    \caption{The calculation of OPE coefficients using the inner product, gluing together two copies of $B^3$ to form a copy of $S^3$ on which we evaluate the inner product itself.}
    \label{fig:opecoeffcalc}
\end{figure}
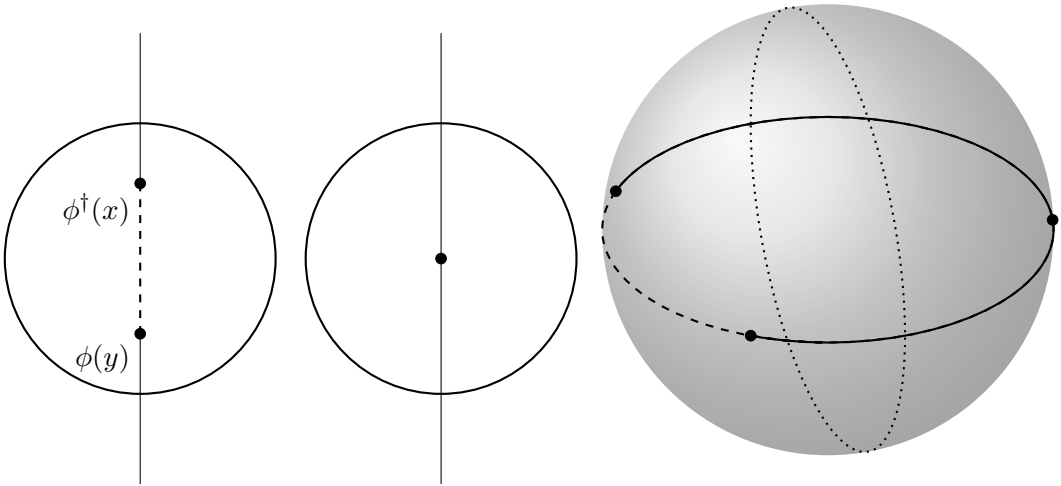

Now let us consider foliating $\mathbb{R}^3$ in radial coordinates with a family of $S^2$'s such that the line defect intersects the sphere at its north and south poles, as in Figure \ref{fig:opecoeffcalc}. It is useful to perform the regular Weyl rescaling to $S^2 \times \mathbb{R}$:
\be
dr^2 + r^2 d\Om^2 = r^2\le(d\tau^2 + d\Om^2\ri) \qquad r = e^{\tau} \label{Weyl} 
\ee
In this conformal frame we may view the theory as defined by a Hamiltonian $H$ that generates $\tau$ translations. The system lives on  $S^2$ spatial slices, and the presence of the defect manifests itself as a deformation of the Hamiltonian -- we will refer to it as a puncture -- at the north and south poles of the $S^2$ $\th = 0, \pi$, thus breaking the $SO(3)$ rotational symmetry down to $SO(2)$. 

We can now construct the space of states on this double-punctured sphere. Directly analogously to the regular state-operator correspondence, each state on the sphere in the $S^2 \times \mathbb{R}$ conformal frame now corresponds to a local operator localized on the line defect in the $\mathbb{R}^3$ conformal frame. Further the energy $\Delta$ with respect to the dilatation operator $H$ corresponds to the defect conformal dimension. 

We now turn to a defect with endpoint, as in \eqref{one-end-point}. Consider placing the endpoint at the origin $x = 0$ and then perform the Weyl rescaling \eqref{Weyl}. We then find a theory living on $S^2 \times \mathbb{R}$ where the Hamiltonian $H$ defining the theory now has only a {\it single puncture} at the north pole of the $S^2$ \cite{Zhou:2023fqu}. Via the same state-operator correspondence, the states $\ket\psi_{\al}$ living on this single-punctured sphere are now in one-one correspondence with different ways to terminate the line defect. These states can be organized into representations of the $SO(2,1)$ conformal algebra and thus transform as defect conformal primaries. Expectation values $\vev{W(x,y)_{\al\beta}}$ as in \eqref{with-infinity} can be related to inner products $\langle \psi_{\beta} | e^{-H\Delta\tau} | \psi_{\al}\rangle$ where in a slight abuse of notation $e^{-H\Delta\tau}$ is  an appropriate element of the defect conformal group $SO(2,1)$ which depends on the insertion points $x$, $y$. 

This state-operator correspondence demonstrates the existence of a defect OPE that allows us to glue together two line defects with endpoints into a single unbroken line defect. We take as an example two Wilson lines with endpoint \eqref{one-end-point}.
One might wonder whether it is possible to glue together two of these Wilson lines to obtain a single unbroken Wilson line $W(C)$, possibly decorated with defect operators $\sO_{\Delta}$ localized at the junction:
\be
W_{\al}(x,\infty)W^{\dagger}_{\beta}(x+\ep,\infty) = W(C)  \sum_{\Delta} c_{\al\beta \Delta} \sO_{\Delta}(x) \ep^{-2\Delta_{\phi} + \Delta} \label{gluingOPE} 
\ee
(Here we ignore possible physics at infinity, assuming e.g. that the line is unbroken there). The state operator correspondence above states that this is possible, following the standard reasoning for bulk conformal field theories (see e.g. \cite{Rychkov:2016iqz} for a review and \cite{Pappadopulo:2012jk} for a detailed argument). To understand this, consider surrounding both of the endpoints of the defect with an $S^2$ as shown in Figure \ref{fig:opecoeffcalc}. This defines a state $\ket{\psi_{\al\beta}}$ on the $S^2$. This will generically not be an eigenstate of the dilatation operator $H$. However it can be decomposed into a sum of eigenstates $\ket{\Delta}$ under $H$
\be
\ket{\psi_{\al \beta}} = \sum_{\Delta} c_{\al\beta \Delta}  \ket{\Delta} \label{decomp} 
\ee
By the correspondence above, each of these states can be mapped to a local operator on the unbroken defect line with a given scaling dimension $\Delta$, resulting in \eqref{gluingOPE}. 

This also provides us with an efficient way to compute the OPE coefficients. From \eqref{decomp} the inner products required $c_{\al\beta \Delta} = \langle \Delta | \psi_{\al\beta} \rangle$ can be computed from the path integral as shown in Figure \ref{fig:opecoeffcalc}. In more detail: the state $\ket{\psi_{\al \beta}}$ is prepared on the left-hand panel by surrounding both endpoints with an $S^2$ and performing a path integral over the interior (topologically a $B^3$), and the state $\ket{\Delta}$ is prepared a simliar way by surrounding the insertion of a single defect-localized operator $\sO_{\Delta}$ on an unbroken Wilson line with an $S^2$ and performing the same interior path integral. The inner product is computed by gluing these two states together along the equatorial $S^2$ to obtain a 3-point function of defect-localized operators (two of which are endpoints) on an $S^3$. The 3-point function of defect-localized conformal primaries is completely determined by conformal kinematics up to an overall prefactor, which determines $c_{\al\beta\Delta}$ in \eqref{decomp}. As usual for an OPE, we expect this to converge provided no other external operators are brought closer than the separation distance $\ep$. 

%We may see from Fig. \ref{fig:opecoeffcalc} that one can calculate the defect three-point function on $S^3$ as shown in the final diagram. This is done by stitching together two copies of $B^2$ onto $S^3$. The importance of this method is that it shows OPE coefficients are obtainable even in the defect case by carefully matching punctured spheres along their boundaries. 

In what follows we will work out some simple  implications of these ideas for the physically interesting theory of large $N$ QED$_3$.

\section{Wilson lines with no end}\label{sec:wilsonnoend}

%In the theory with a Wilson line inserted spanning the $z$-axis $\{\textbf{r} \in \mathbb{R}^3 : r_x=r_y=0\}$, note that this set is mapped to $\{(\theta=0), (\theta=\pi)\}$ on $S^2 \times \mathbb{R}$. Thus, we make use of a state-operator correspondence for the theory with the Wilson line defect, in which states are formulated on a punctured sphere $S^2 \setminus \{\theta=0, \theta=\pi\}$. 

%We will be interested in writing down the OPE coefficients defined in the introduction to leading order in $1/N_b$. At this order, as we shall see in subsequent calculations, the scalar fields at the endpoints of the Wilson lines are free scalars each with dimension $1/2$, and the defect-localised scalar, as derived earlier, has dimension $n+m+1/2$. In particular, the lowest energy state on $S^2 \times \mathbb{R}$ corresponds to a scalar with dimension $1/2$ (when $n=j=m=0$).

%This means that the final quantity of interest is a defect three-point function on $S^3$, and in each case, the OPE coefficient is a number prefacing the positional part of these functions. We shall illustrate the calculation of a few of these simple quantities in the final section.

We first study the theory defined by \eqref{CPnR3} in the presence of a single infinite Wilson line with charge $q$ on $\mathbb{R}^{3}$, extending in the $z$ direction along $x,y = 0$:
\be
W(C) \equiv \exp\le(iq\int_{-\infty}^{+\infty} dz A_{z}\ri) \label{wilsoninf} 
\ee
The presence of this Wilson line will create a non-trivial field strength for the dynamical gauge field $A$ and Hubbard-Stratonovich field $\lam$. By conformal invariance, these fields on $\mathbb{R}^3$ are determined up to overall prefactors as
\begin{equation}
F^{\mathbb{R}^3}_{iz} = \frac{iE x_i}{\rho^3} \qquad \lambda^{\mathbb{R}^3} = \frac{\lambda_0}{\rho^2} \qquad \rho = \sqrt{x^2 + y^2} \label{Fprof} 
\end{equation}
We note that the conformal dimension of $F$ is fixed at $2$ for all $N$ (as it is the conserved current for the topological symmetry $U(1)_m$) but the dimension of $\lam$ is in general modified from its tree-level value away from infinite $N$ and thus this calculation is valid only to leading order at large $N$. 

\subsection{Coefficient of electric field} 
The values of $E$ and $\lam_0$ can be computed in terms of the charge of the Wilson line $q$ at large $N$. We begin by presenting a simple calculation for the value of $E$. In particular, at the conformal fixed point the bare kinetic term $F^2$ for the gauge field is irrelevant and can be omitted; instead its dynamics can be obtained by integrating out the bosonic fields in the $1/N$ expansion. The quadratic effective action thus obtained is:
\begin{equation}\label{eq:effective_double_puncture_action}
S_{eff} [A] = \text{const} + \frac{N}{32} \int \frac{d^3 k}{(2\pi)^3} |k| \left(\delta^{\mu \nu} - \frac{k^\mu k^\nu}{k^2} \right) A_\mu(k) A_\nu(-k) + ... 
\end{equation}
For a review of the derivation of this standard effective action, see Appendix \ref{app:deriv_effective_action}. We now consider varying the action in the presence of the Wilson line source $S_{eff}[A] + i q\int dz A_{z}$. The equation of motion \textit{in momentum space} to leading order is
\begin{equation}
\frac{N_b}{16} A_\nu(k) |k| \left(\delta^{\mu \nu} - \frac{k^\mu k^\nu}{k^2}\right) = 2i\pi q \delta^{z \mu} \delta(k_z) \label{sourceeqn} 
\end{equation}
%(Note the 8 is because of a factor of 2 that comes from differentiating $A^2$!)
Here we have on the right hand side taken the Fourier transform of the usual right hand side of the equation of motion in \textit{position space}, which is $q \delta^{z \mu} \delta^{(2)} (x,y)$. The Fourier transform of the saddle-point $A_\mu$ associated with \eqref{Fprof} 
is:
\begin{equation}
A_\mu(k) = i E \delta^z_\mu \int d^3 x \, e^{-i\textbf{k}\cdot\textbf{x}} \frac{1}{\sqrt{x^2+y^2}} = 4\pi^2 i E \frac{\delta^z_\mu \delta(k_z)}{\sqrt{k_x^2+k_y^2}}
\end{equation}
Where $E$ is at this stage an undetermined constant to be determined by plugging into the equation of motion. We can do this now, obtaining:
\begin{equation}
\frac{N \pi^2}{4} E \frac{|k|\delta(k_z)}{\sqrt{k_x^2+k_y^2}} \left(\delta^{\mu z} - \frac{k^\mu k^z}{k^2}\right) = 2q\pi \delta^{z \mu} \delta(k_z)
\end{equation}
We can see that this equation holds because of the $\delta(k_z)$: once we transform back to position space, it kills the second term, and we have:
\begin{equation}
E= \frac{8 q}{N \pi} \label{eval} 
\end{equation}
where this calculation is correct to leading order in $N^{-1}$. 

Calculating the saddle-point value of $\lam_0$ is somewhat more intricate as it begins at order $N^{-2}$, and for this we need to perform a more explicit one-loop calculation. 
%In the language of our earlier calculations, $\alpha=E$, so that $A = iE / \sqrt{x^2+y^2}$, as we used in our earlier formulae concerning states on the double-punctured sphere. 
%We will now quantize the theory on $S^2 \times \mathbb{R}$ and understand the spectrum of the theory. 

\subsection{Spectrum and one-loop calculation} \label{infspect} 

%In this section, we focus on studying the stability of Wilson lines with no ends. We pick, as described before, an infinite Wilson line extending along the $z$-axis, so as to break the conformal symmetry in a controlled fashion to $SO(2) \times SL(2;\mathbb{C})$. To be precise, we are deforming the theory by the insertion of an operator
%\begin{equation}
%W(\{(0,0,z) | \, z \in \mathbb{R}\}) = \exp \left(\int_{-\infty}^{\infty} dz \,  A_z\right)
%\end{equation}

%First, we need to find the saddles of $F$ and of $\lambda$ in order to use these in the scalar equation of motion to find the states we desire. In \cite{Aharony:1998qu}, it is argued from conformal symmetry considerations that around a Wilson line extending infinitely along the $z$-axis, the saddle-point field strength must take the form
%\begin{equation}
%F_{iz} = \frac{iE x_i}{\rho^3}
%\end{equation}
%Where $\rho = \sqrt{x^2+y^2}$. The constant $E$ will be derived in terms of the charge $q$ of the Wilson line in the following section of the paper. It will be convenient to transform this to $S^2 \times \mathbb{R}$ co-ordinates:
%\begin{equation}
%F_{\tau \theta} = - \frac{iE}{\sin^2 \theta} 
%\end{equation}
%which can be written as $F=d\mathcal{A}$ with 
%\begin{equation}
%\mathcal{A} = iE \cot \theta d\tau
%\end{equation}
%in a particularly simple choice of gauge. We will take this saddle-point strength as the leading order contribution in the scalar field equation of motion.

We will now quantize this theory on $S^2 \times \mathbb{R}$ written as
\be
ds^2 = d\tau^2 + d\theta^2 + \sin^2 \th d\phi^2
\ee
As described earlier, the insertion of the Wilson line \eqref{wilsoninf} now corresponds to the deformation of the $S^2$ Hamiltonian by two explicit insertions of the gauge field at $\th = 0, \pi$:
\be
H_{\mathrm{Wilson}} = H_{\mathrm{CFT}} + iq (A_{\tau}(\th=0) - A_{\tau}(\th = \pi))
\ee
We can obtain the conformally invariant form of the background fields by performing the appropriate change of variables and Weyl rescaling on \eqref{Fprof}:
\be
F^{S^2 \times \mathbb{R}}_{\tau \theta} = - \frac{iE}{\sin^2 \theta} \qquad \lambda^{S^2 \times \mathbb{R}} = \frac{\lambda_0}{\sin^2 \theta}
\ee
In the remainder of this subsection we will work only in the $S^2 \times \mathbb{R}$ conformal frame and so omit the superscripts. We choose the following gauge:
\be
A = iE \cot \theta d\tau
\end{equation}
One can check explicitly that this profile is invariant under $SO(2,1)$ up to gauge transformations. 

From this point, we shall calculate several quantities whose gauge-invariance or variance we shall briefly make explicit here. The saddle profile of $A_\mu$ that shall be used in calculations, presented above, is evidently gauge-dependent. With this profile we calculate the fluctuation spectrum on the Wilson line, which is gauge-dependent level-by-level. However, the ground-state energy on $S^2 \times \mathbb{R}$ calculated is gauge-invariant. The conditions for marginality and instability of the Wilson line we shall establish, further, are also gauge-invariant. Finally, in Section IV, we will consider a Wilson line ending at a finite point and compute the one-point function of the field strength $F_{\mu\nu}$, which is gauge-invariant.

We will now compute the one-loop free energy, minimizing it with respect to $\lam_0$ to determine its saddle-point value. A similar calculation was performed in \cite{Dyer:2015zha} in the presence of a monopole background. Finding the saddle is equivalent to solving the variational equation
\begin{equation}\label{eq:saddlepointlambda}
\frac{\delta S_{eff} \left[A,\lambda\right]}{\delta \lambda} = 0
\end{equation}
where the effective action after integrating out the scalar fields is given by
\begin{equation}\label{eq:effectiveactionwithhsfield}
S_{eff} \left[A,\lambda\right] = N \left(\left[ \tr \log \left( - D_\mu D^\mu + \lambda + \frac{1}{4}\right)\right] - \frac{1}{\frak{g}} \int d^3 x \, \sqrt{g} \lambda \right)
\end{equation}
(Note that we have absorbed a factor of $i$ into the definition of $\lambda$ compared to the equivalent expression in \cite{Dyer:2015zha}).

It is important to clarify what is meant by finding the saddle point of $\lambda$ as in \eqref{eq:saddlepointlambda}.  Following \cite{Dyer:2015zha}, we will not take a derivative in functional space, as conformal invariance dictates that the only free parameter is $\lam_0$; thus we compute the on-shell action as a function of $\lam_0$ and optimize with respect to the value of that parameter.  %However, this is difficult to do rigorously. nIn \cite{Dyer:2015a} the authors first restrict attention to a physically symmetry-constrained set of $\lambda$s, and then find the saddle-point over this set, having already substituted the appropriate form of $\lambda$ into the effective action \eqref{eq:effectiveactionwithhsfield}. We follow the same approach.

The effective action for this ansatz then becomes
\begin{equation}
S_{eff} \left[A,\lambda\right] = N \left(\left[ \tr \log \left( - D_\mu D^\mu + \lambda + \frac{1}{4}\right)\bigg\rvert_{\lambda=\lambda_0 /\sin^2 \theta}\right] - \frac{2\pi\lambda_0}{\frak{g}} \int_0^\pi \frac{d\theta}{\sin \theta} \right)
\end{equation}
It is apparent that the integral on the right-hand side is divergent, yet we shall ignore this for now, and see later that divergences cancel appropriately.

Now, the action of the operator $-D_\mu D^\mu + \lambda + 1/4$ on $\Phi$ can be found from \eqref{scalac} and explicitly written as
\begin{align}
-\frac{1}{\sin^2 \theta} [\partial_\phi^2 \Phi + \sin^2 \theta \partial_\tau^2 \Phi + \sin\theta \partial_\theta (\sin \theta \partial_\theta \Phi) + ... \nonumber \\...+ (E^2 + E \cos \theta \sin \theta) \Phi - 2E \cos \theta \sin \theta \partial_\tau \Phi - \frac{1}{4} \sin^2 \theta \Phi - \lambda_0 \Phi ]
\end{align} 
This operator has eigenfunctions given by
\begin{equation}
\Phi_{n,m,E}(\tau,\theta,\phi) = e^{\Delta \tau} e^{im\phi} (e^{i\theta})^{- n - i\mathcal{E}} (2i\sin \theta)^{+ \tilde{m}} \;_2 F_1 \left(\frac{1}{2} - i\mathcal{E} + \tilde{m},-n,\frac{1}{2}-i\mathcal{E} - n - \tilde{m},e^{2i\theta}\right)
\end{equation} 
where we have defined the following
\begin{equation}
\tilde{m}^2 = m^2+ \lambda_0-E^2
\end{equation}
and where the parameter $\mathcal{E}$ takes the specific value 
\begin{equation}
\mathcal{E}=\frac{2 E \Delta}{1+2n+2\tilde{m}} \ . 
\end{equation}
The corresponding eigenvalues -- found by demanding regularity on the sphere -- are
\begin{equation}
\Lam_{n,m} = \left(\Delta^2 - (n+\frac{1}{2}+\tilde{m})^2\right)\left(1+\frac{E^2}{(n+\frac{1}{2}+\tilde{m})^2}\right)
\end{equation}
where $m \in \mathbb{Z}$ is the regular azimuthal quantum number and $n \in \{0, 1, 2, \cdots \}$ is a discrete quantum number  describing dynamics in the $\th$ direction. 
The on-shell spectrum is found when the equations of motion are satisfied, i.e. when $\Lam_{m,n} = 0$, where positive energies are given by
\begin{equation}
\Delta =  \left(n+\frac{1}{2}+\tilde{m}\right) \qquad \tilde{m}^2 = m^2+ \lambda_0-E^2 \label{deltadef} 
\end{equation}
%It is also important to note that since $E \sim q/N_b$, to leading order in $1/N_b$, the eigenvalues are just
%\begin{equation}
%\left(\Delta^2 - (n+\frac{1}{2}+\tilde{m})^2\right) + \mathcal{O} \left(\frac{1}{N_b^2}\right)
%\end{equation}
%This will reappear when we do the saddle-point calculation explicitly. \NI{okay, I realize that we are actually {\it keeping} the $N^{-2}$ dependence in the computation of $\lam_0$, and thus I think we should {\it not} remove it here. Return to this and carefully organize}. 

Let us study some limits of these expressions. We first orient ourselves by considering the case where $E = \lam_0 = 0$, and so we are simply considering a free scalar with no Wilson line. In the case the spectrum is given by:
\be
\Delta_{\mathrm{free}} = n + |m| + \ha  \label{freespec} 
\ee
In particular, at $n = m = 0$ we find $\Delta = \ha$, which via the state-operator correspondence is indeed the conformal dimension of a free scalar $\phi(x)$ in $3$ dimensions. Excited states carrying angular momentum will correspond to derivatives of $\phi$. It is also instructive to note this can be written in terms of the usual magnitude of the angular momentum $j$ 
\be
\Delta_{\mathrm{free}} = \sqrt{j(j+1) + \frac{1}{4}} = j+\ha \qquad j = n + |m| 
\ee
where we see that the usual degeneracy $2j+1$ is obtained by noting that for a given value of $j = n + |m|$ there are $2j + 1$ values of $m, n$ with the same value of $n + |m|$. 
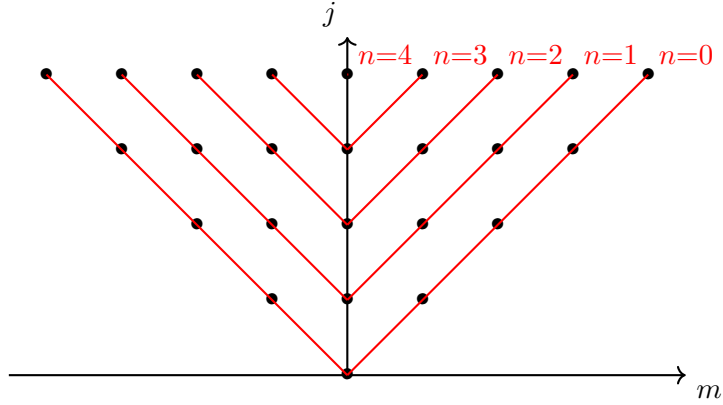
\begin{figure}
    \centering
    \begin{tikzpicture}
        \draw[thick,->] (-4.5,0) -- (4.5,0) node[anchor=north west] {$m$};
        \draw[thick,->] (0,0) -- (0,4.5) node[anchor=south east] {$j$};
        \foreach \Point in {(0,0), (1,1), (0,1), (-1,1), (2,2),(1,2),(0,2),(-1,2),(-2,2),(3,3),(2,3),(1,3),(0,3),(-1,3),(-2,3),(-3,3),(4,4),(3,4),(2,4),(1,4),(0,4),(-1,4),(-2,4),(-3,4),(-4,4)}{
            \node at \Point {\textbullet};
        }
        \draw[red,thick,-] (0,0) -- (4,4) node[anchor=south west] {$n$=0};
        \draw[red,thick,-] (0,0) -- (-4,4);
        \draw[red,thick,-] (0,1) -- (3,4) node[anchor=south west] {$n$=1};
        \draw[red,thick,-] (0,1) -- (-3,4);
        \draw[red,thick,-] (0,2) -- (2,4) node[anchor=south west] {$n$=2};
        \draw[red,thick,-] (0,2) -- (-2,4);
        \draw[red,thick,-] (0,3) -- (1,4) node[anchor=south west] {$n$=3};
        \draw[red,thick,-] (0,3) -- (-1,4);
        \draw[red,thick,-] (0,4) -- (0,4) node[anchor=south west] {$n$=4};
    \end{tikzpicture}
    \caption{Summing quantum numbers $n=j-|m|$ and $m$ in the $E=0$ case}
    \label{fig:quantumnumbers}
\end{figure}

Let us now also consider the case of the tri-critical model, where we recall from the discussion around \eqref{CPnR3} that we have $\lam_0 = 0$, and thus the spectrum is simply
\be
\Delta_{\mathrm{tri-critical}} = n + \sqrt{m^2 - E^2} + \ha
\ee
where we still have $E \sim \frac{q}{N}$ from \eqref{eval}. Note in particular that the state with $m = 0$ is always unstable for any arbitrarily small $E$; this shows that Wilson line always has an instability, as argued for on slightly different grounds in \cite{Aharony:2023amq}. Physically this can be traced back to the divergence of the gauge potential at the poles, which creates a pair-creation instability. Modes that are smooth on the sphere behave near the pole as $(r e^{i\phi})^m$ where $r$ is the distance from the pole, and thus modes with $m > 0$ vanish at the pole and are insensitive to the divergence; however the modes with $m = 0$ have nontrivial support near the pole and become unstable in the presence of the diverging electric field. 

We turn now to our main interest, the $\mathbb{CP}^{N-1}$ model. Here the profile of $\lam$ -- which acts like a $\th$-dependent mass -- has the possibility to cure the instability provided its saddle-point value is large enough. We now compute this saddle-point value by varying over $\lam_0$. 

Inserting the expression for the eigenvalues we find:
\begin{flalign}
&\mathcal{F} [A, \lambda]\bigg\rvert_{\lambda=\lambda_0/\sin^2 \theta} \nonumber \\ &=  \int \frac{d\omega}{2\pi} \left( \sum_{m\in \mathbb{Z}, n \in \mathbb{Z}_{\geq 0}} \log \left(\left(\omega^2 + (n+\frac{1}{2}+\tilde{m})^2\right)\left(1+\frac{E^2}{(n+\frac{1}{2}+\tilde{m})^2}\right)\right) \right) - \frac{2\pi\lambda_0}{\frak{g}} \int_0^\pi \frac{d\theta}{\sin \theta}
\end{flalign}
Where $S_{eff}[A,\lambda] = N \mathcal{F} [A, \lambda]$. For brevity let us write
\begin{equation}
\mathcal{F}_1[A,\lambda_0] =  \int \frac{d\omega}{2\pi} \left( \sum_{m\in \mathbb{Z}, n \in \mathbb{Z}_{\geq 0}} \log \left(\left(\omega^2 + (n+\frac{1}{2}+\tilde{m})^2\right)\left(1+\frac{E^2}{(n+\frac{1}{2}+\tilde{m})^2}\right)\right) \right)
\end{equation} 
To leading order in $1/N$, our saddle-point condition becomes% (when we tune the coupling to the critical coupling $\frak{g}_c$, defined shortly)
\begin{equation}\label{eq:fsaddlecondition}
\frac{\partial \mathcal{F} \left[A,\lambda_0\right]}{\partial \lambda_0} = 0
\end{equation}
Note however that we also have to make sure that the theory is tuned to criticality in the absence of any Wilson lines. As in \cite{Dyer:2015zha}, this is done by setting the coupling $\frak{g}$ to the critical coupling $\frak{g}_c$. This critical coupling is defined by demanding that in the absence of a Wilson line, i.e. $E = 0$, the scalar is conformally coupled, i.e. the saddle point value of $\lam_0 = 0$. In other words, the definition of $\frak{g}_c$ is:
\begin{equation}
\frac{\partial \mathcal{F}_1 \left[A,\lambda_0\right]}{\partial \lambda_0} \bigg\rvert_{\lambda_0=E=0} - \frac{2\pi}{\frak{g_c}} \int_0^\pi \frac{d\theta}{\sin \theta} = 0
\end{equation}
Thus, imposing $\mathfrak{g} = \mathfrak{g}_c$, at criticality our saddle-point condition \eqref{eq:fsaddlecondition} now becomes
\begin{equation}
\frac{\partial \mathcal{F}_1 \left[A,\lambda_0\right]}{\partial \lambda_0} - \frac{\partial \mathcal{F}_1 \left[A,\lambda_0\right]}{\partial \lambda_0} \bigg\rvert_{\lambda_0=E=0} = 0.
\end{equation}
By construction $E=0$, $\lambda_0=0$ satisfies the above saddle-point condition. 
 
To compute the sum over eigenvalues above, we note that standard manipulations of the integration contour (reviewed in Appendix \ref{app:onshell}) can be used to perform the integration over $\om$ through residues, eventually obtaining the answer in terms of a sum over the on-shell frequencies $\om= i\Delta_{n,m}$:
\be
\frac{\p \sF_1[A,\lam_0]} {\p \lam_0} = \sum_{m,n} \frac{\p \Delta_{m,n}(\lam_0)}{\p \lam_0} \label{contint} 
\ee
whose definition from \eqref{deltadef} we repeat here:
\be
\Delta_{n,m} =  \left(n+\frac{1}{2}+\sqrt{m^2 + \lam_0 - E^2}\right)
\ee
Differentiating with respect to $\lam_0$ we find that the full saddle-point condition is:
\begin{equation}
\sum_{m\in \mathbb{Z}, n \in \mathbb{Z}_{\geq 0}} \left( \frac{1}{2 \tilde{m}} -  \frac{1}{2 m} \right) = 0.
\end{equation}
Interestingly, one need not perform any sums to solve this equation for $\lam_0$; from the definition of $\tilde{m}$ in terms of $m$ we see that this equation is immediately satisfied if $\lam_0 = E^2$. Physically, setting the Hubbard-Stratonovich saddle to this value cures the instability seen in the tricritical case, where $\lambda_0 = 0$. 

Note that though these computations have been carried out in a particular choice of gauge for $A$, and the specific scalar field profile is gauge-dependent, the conclusion of stability in this case is a gauge-invariant one, since changing gauge yields a  different operator whose eigenfunctions are the same as those we have written out above just combined with a phase shift, and the trace log of the operator therefore remains identical. The statement on stability is therefore physical.

Interestingly, with this on-shell value of $\lam_0$ the Hubbard-Stratonovich field completely screens the electric field. Thus, not only is the Wilson line stable, but also the spectrum of single-particle defect operators is the same as though there were no Wilson lines present: 
\be
\Delta_{\mathbb{CP}^{N-1}} = n + |m| + \ha
\ee
This somewhat surprising conclusion about the scalar field states is valid for an arbitrary background $E$. However our solution to the dynamical gauge field equation to relate the background $E$ to $\frac{q}{N}$ is itself only correct to linear order in $N^{-1}$, and we expect $N^{-1}$ quantum corrections to the scalar field to result in the spectrum being sensitive to the Wilson line.

We may now consider the spectrum of defect-localized states created by the $\phi$ field; we can canonically quantize 
\be
\phi^{\al}(\tau, \th, \phi) = \sum_{m,n} \le(e^{i\om_{m,n} \tau} a^{\al}_{m,n} f_{m,n}(\th,\phi) + e^{-i\om_{m,n} \tau} (b_{m,n}^{\al})^{\dagger}f_{m,n}^{\dagger}(\th,\phi)\ri)
\ee
and act on the vacuum (i.e. the undecorated defect) with particle and anti-particle creation operators. Note that the net $U(1)$ gauge charge must be zero due to the equation of motion for $A_{\tau}$, and thus each state in the physical subspace has equal numbers of particles and antiparticles, corresponding to gauge-invariant products of $\phi \phi^{\dagger}$. At this order in $N^{-1}$ the energies of the scalars simply add, though one could compute corrections systematically in powers of $\sO(N^{-1})$. 

We note that when a similar analysis is carried out for the Chern-Simons theory in \cite{Gabai:2022vri}, a distinct stable/unstable structure is found for different values of coupling at the endpoint. No direct analogue of such a structure seems to exist for our setup.

%\NI{We should actually discuss what happens to the singlet operator that was offensive in the tri-critical model}

A deficiency of our analysis is that the relation between $E$ and $q$ in \eqref{eval} was obtained only in a linearized analysis about $E = 0$. In principle one can obtain a non-linear action for $E$ through a one-loop determinant calculation similar to the one we just performed for $\lam$. In the $S^2$ conformal frame that we use this computation is plagued by divergences at the poles of the sphere, and we have not pursued this here. We note however that a similar computation was carried out for fermions using an AdS$_2$ conformal frame in \cite{Aharony:2023amq}, where the presence of a regulator that exploited the volume of AdS$_2$ allowed divergences to be easily regulated; we hope to return to this in the future. 
 
\section{Wilson lines with endpoints} \label{sec:wilsonend} 
We now turn to a Wilson line with endpoints. As described around \eqref{with-infinity}, the simplest Wilson line with endpoints takes the form
\be
W(x_1,x_2)_{\al\beta} = \phi_{\beta}^{\dagger}(x_2)\exp\le(i\int_{x_1}^{x_2} A\ri) \phi_{\al}(x_1) \label{repeatWilson} 
\ee
Though one is tempted to place one of the endpoints at infinity and thus ignore it, in a conformal theory this option is not available, as infinity is just another point; thus our notation discusses the location of each endpoint explicitly. We will first determine the saddle-point value of $\langle F_{\mu\nu}(x)\rangle_{W}$ in the presence of the defect. Formally this is a three-point function of two defect-localized operators $\phi_{\al}(x_1), \phi_{\beta}(x_2)$ and one bulk operator $F_{\mu\nu}(x)$:
\be
\langle F_{\mu\nu}(x)\rangle_{W}= \frac{\langle F_{\mu\nu}(x) \phi_{\beta}^{\dagger}(x_2)e^{i\int_{x_1}^{x_2} A}  \phi_{\al}(x_1) \rangle}{\langle \phi_{\beta}^{\dagger}(x_2)e^{i\int_{x_1}^{x_2} A} \phi_{\al}(x_1)\rangle} \label{Fdef3pt} 
\ee
where we have normalized it appropriately. Note that the spatial profile of such a 3-point function -- unlike that of three defect-localized operators -- is not fixed by conformal invariance and will thus differ from theory to theory. We now calculate it explicitly in QED$_3$ to leading order in $N^{-1}$. As the base object $\vev{W_{\al\beta}}$ vanishes unless $\al = \beta$, from now on we take $\al = \beta$ and it is understood that there is no sum on $\al$. 

%We have already introduced this object earlier, and write it, following notation consistent throughout the paper, as
%\begin{equation}
%W((x,\infty);x) = \exp\le(i\int_{x}^\infty A\ri) %\phi(x)
%\end{equation}
%It is clear that a gauge transformation $\phi(x) \rightarrow e^{i\alpha(x)} \phi(x)$, $A_\mu \rightarrow A_\mu + \partial_\mu \alpha$ leaves this object unchanged. In what follows, we shall consider first an infinite Wilson line extending along the z-axis, and then two Wilson lines with endpoints, along the z-axis, where these endpoints are separated by some small distance in terms of which we will eventually formulate an OPE.

\subsection{Saddles in $F_{\mu\nu}$}

We will compute this saddle point value as we did for the case of the infinite Wilson line, i.e. by expanding around $A=0$ in the $1/N$ expansion with the effective action:
\begin{equation}\label{eq:effective_double_puncture_action}
S_{eff} [A] = \text{const} + \frac{N}{32} \int \frac{d^3 k}{(2\pi)^3} |k| \left(\delta^{\mu \nu} - \frac{k^\mu k^\nu}{k^2} \right) A_\mu(k) A_\nu(-k) + ...
\end{equation}
We now place our Wilson line extending in the $z$ direction in a straight line between the two points $x_1 = (0,0,0)$ to $x_2 = (0,0,\infty)$. As the defect only extends for the half-line $z \in [0,\infty)$, one is tempted to consider the action $S_{eff}[A] + i\int_0^{\infty}dz A_{z}$. Varying this action we would find an expression similar to \eqref{sourceeqn}, except that the source on the right-hand side would be the Fourier transform of the \textit{new} half-line defect source:
\begin{equation}
i \delta^{z \mu} \delta^{(2)} (x,y) \Theta(z) 
\end{equation}
where $\Theta$ is the Heaviside step function, corresponding to the Wilson line with an endpoint at the origin, rather than the full infinite Wilson line. This expression contains no $q$ since we will be considering a Wilson line with a single quantum of charge.

The equation of motion that is the analogue of the extended Wilson line in \eqref{sourceeqn} is
\begin{equation}
\frac{N}{16} A_\nu(k) |k| \left(\delta^{\mu \nu} - \frac{k^\mu k^\nu}{k^2}\right) \stackrel{?}{=} i\delta^{z\mu} \left(\pi \delta(k_z) - \frac{i}{k_z}\right) \label{sourceeqn} 
\end{equation}
where the $k_z^{-1}$ arises from the endpoint; we outline the derivation in Appendix. 
However this equation cannot be correct. By construction the left-hand side can only result in an equation in Fourier space that is orthogonal to $k^\mu$, which this new half-line source is not. Equivalently, we may see that in position space the source is no longer divergenceless. 

All of this is a manifestation of the fact that the half-Wilson line without the scalar field attached is not gauge invariant. To properly account for the effect of this endpoint of the Wilson line, we must include the scalar field $\phi_a(0)$ attached there within our calculation of $S_{eff}[A]$. 

First, note that to write the action succinctly, we will make use of the global current $j^\mu$ that couples to the gauge field $A_\mu$:
\begin{equation}
j^\mu_l = i(\phi^\dag_l \partial^\mu \phi_l - \phi_l \partial^\mu \phi^\dag_l) \label{jdef} 
\end{equation}

Following exactly the same procedure as the full Wilson line case that is detailed in Appendix \ref{app:deriv_effective_action}, we write the total action of the gauge field $A$ found by integrating out the scalars as 
\begin{flalign}
e^{-S_{tot}[A]} &= \frac{1}{Z_\phi} \lim_{w\rightarrow \infty}\int \prod_{i=1}^{N_b} \mathcal{D}\phi_i \, \frac{\phi_\al(0) \phi^\dag_\al(w)}{\langle \phi_\al(0) \phi^\dag_\al(w) \rangle} e^{\left(-\sum_{l} \int d^3 x \,  j_l^\mu(x)A_\mu(x) +  A^2(x) \phi_l^\dag \phi_l(x)\right)-\sum_{j} \int d^3 x \, \mathcal{L}(\phi_j(x))} e^{\int_0^\infty dz A_z(z)} \\ 
\Rightarrow S_{tot}[A] &= -\lim_{w\rightarrow \infty} \int d^3 x \,  \left(\delta^{(2)}(x,y) \Theta(z) A_z(x) - \frac{\langle \phi_\al(0) j^\mu_\al(x) \phi^\dag_\al(w) \rangle}{\langle \phi_\al(0) \phi^\dag_\al(w) \rangle} A_\mu(x) \right) + \text{const.} \nonumber \\  - \frac{1}{2}\sum_{\ga,\delta}\int &d^3 x \int d^3 x' \, \langle j_\ga^\mu(x) j_\delta^\nu(x') \rangle A_\mu(x) A_\nu(x') +...
\end{flalign}
Again, the expectation value is in the undeformed CFT,  $\langle \, \cdot \, \rangle = \langle \, \cdot \, \rangle_{A=0} $. The last term here leads to the effective action we have already been using.

We keep the terms above because $A$ admits an expansion in powers of $1/N_b$, and we will keep the leading term, in the form:
\begin{equation}
A_\mu(k) = \frac{1}{N} A^0_\mu(k) + \mathcal{O} \left(\frac{1}{N^2}\right)
\end{equation}
To find the profile $A^0_\mu(k)$, we work in momentum space. Recall that in 3D we have the propagator: 
\begin{equation}
\langle \phi_{\al}(x) \phi_{\beta}(y) \rangle = \delta_{\al\beta} G(x,y) \qquad G(x,y) = \frac{1}{4\pi} \frac{1}{|x-y|}
\end{equation}
Using the expression for $j_a^{\mu}$ \eqref{jdef} we  immediately note that many of the terms above are therefore subleading: note that
\begin{equation}
\lim_{w \rightarrow \infty} \frac{\frac{\partial}{\partial x^\mu} G(x,w) }{G(0,w)} = \lim_{w \rightarrow \infty} \frac{w^\mu}{|w|^2} + ... = 0 +...
\end{equation}
while
\begin{equation}
\lim_{w \rightarrow \infty} \frac{G(x,w) }{G(0,w)} = \lim_{w \rightarrow \infty} 1 + \frac{x \cdot \hat{w}}{|w|} +... = 1 + ...
\end{equation}
Therefore we can drop Wick contractions between the scalar field at $x$ and at $\infty$ when there is a derivative involved. So only one term contributes, and this gives:
\begin{equation}
\frac{\langle \phi_{\al}(0) j^\mu_{\al}(x) \phi^\dag_{\al}(w) \rangle}{\langle \phi_{\al}(0) \phi^\dag_{\al}(w) \rangle} = \frac{i\langle \phi_{\al}(0) (\phi^\dag_{\al}(x) \partial^\mu \phi_{\al}(x) - \phi_{\al}(x) \partial^\mu \phi^\dag_{\al}(x)) \phi^\dag_{\al}(w) \rangle}{\langle \phi_{\al}(0) \phi^\dag_{\al}(w) \rangle} \to \frac{i}{4\pi} \frac{x^\mu}{|x|^3}
\end{equation}
where in the last limit we take $w \to \infty$. 
To Fourier transform this, we'll make use of the formula:
\begin{equation}\label{eq:skenderis_formula}
\int d^3 x \, \frac{e^{-ik \cdot x}}{|x|^{2n}} = \frac{\pi^{3/2} 2^{3-2n} \Gamma \left(\frac{3}{2} - n \right)}{\Gamma(n)} |k|^{2n - 3}
\end{equation}
In particular, when $n=1/2$, we get $4\pi/|k|^2$. So the equation of motion now to leading order in $1/N_b$ is:
\begin{equation}
\frac{i k^\mu}{|k|^2} + \frac{1}{16} A^0_\nu(k) |k| \left(\delta^{\mu \nu} - \frac{k^\mu k^\nu}{|k|^2}\right) = i\pi \delta^{z \mu} \left(\delta(k_z) - \frac{i}{\pi k_z}\right)
\end{equation}
(Note now that $q=1$ explicitly since we have a \textit{single} charge at the endpoint of the Wilson line). The following ansatz for $A^0_\mu(k)$ solves this equation of motion:
\begin{equation}\label{eq:adjusted_ansatz}
A^0_\mu(k) = \beta \frac{\delta_{z\mu} \delta(k_z)}{\sqrt{k_x^2+k_y^2}} + \frac{i \alpha \delta_{z\mu}}{\pi |k| k_z}
\end{equation}
  Note here that to recover the second term in position space, we can observe that the $z$-derivative of this term in positional space is proportional to the function whose momentum-space representation is $-\alpha \delta_{z\mu} / \pi |k|$. This means that we can write the second term in position space as
\begin{equation}
-\frac{\alpha \delta_{z \mu} \arctan \left(\frac{\sqrt{x^2+y^2}}{z}\right)}{2\pi^3 \sqrt{x^2+y^2}} + \delta_{z\mu}f(x,y)
\end{equation}
for some function $f(x,y)$ of integration, such that the relevant boundary condition here is satisfied: namely, that as $\theta \rightarrow \pi$ in the $S^2 \times \mathbb{R}$ picture, $A_\mu \rightarrow 0$. 

Plugging \eqref{eq:adjusted_ansatz} into the equation of motion, we satisfy it if 
\begin{eqnarray}
\beta = 16i\pi \\
\alpha = -16i\pi
\end{eqnarray}
Transforming this back into position space, we have:
\begin{equation}
A_\mu = \delta_{z\mu} \left(\frac{4}{\pi\sqrt{x^2+y^2}} - \frac{8}{\pi^2\sqrt{x^2+y^2}} \arctan \left(\frac{\sqrt{x^2+y^2}}{z}\right) 
+f(x,y)\right)
\end{equation}
We now impose the boundary condition that as $\theta \rightarrow \pi$, $A_\mu^0 \rightarrow 0$: this is equivalent to picking the choice $f(x,y)=-\frac{4}{\pi \sqrt{x^2+y^2}}$. With this established, the gauge field saddle-point profile becomes
\begin{equation}
A_\mu = \delta_{z\mu} \frac{8i}{N \pi} \left(\frac{1-\frac{1}{\pi}\arctan \left(\frac{\sqrt{x^2+y^2}}{z}\right)}{\sqrt{x^2+y^2}}\right) \label{Aans} 
\end{equation}

From this, the field strength tensor can be written
\begin{equation}
F_{iz} = \frac{8ix^i}{N\pi(x^2+y^2)^{3/2}} \left(\frac{1}{\pi}\arctan \left(\frac{\sqrt{x^2+y^2}}{z}\right)-1-\frac{z\sqrt{x^2+y^2}}{(x^2+y^2+z^2)} \right)
\end{equation} where here $i=x,y$.

Despite the length of the computation the final answer is quite intuitive; as we go further away from the origin, it simply smoothly interpolates between the result for the infinite Wilson line (at positive $z$) and zero (at negative $z$). 

To make the limits explicit, we can write these Cartesian components of the field-strength tensor in terms of polar co-ordinates in $\mathbb{R}^3$ (for the avoidance of confusion, \textit{not} in the $S^2 \times \mathbb{R}$ picture!). 
\begin{equation}
F_{xz} = \frac{8i \cos \phi}{N\pi r^2 \sin^2 \theta} \left(\frac{\theta}{\pi}-1- \sin \theta \cos \theta \right)
\end{equation}
Where $r_\perp = \sqrt{x^2+y^2}$ is the perpendicular distance from the Wilson line.

Consider first the behaviour of this profile far away from the endpoint (to remind: at the origin) in the direction away from the Wilson line, where $z<0$. Here, $\theta \rightarrow \pi$, and by the choice of our earlier function $f(x,y)$, we have $F_{xz} \rightarrow 0$. 

At the other end, as $\theta \rightarrow 0$, $F_{xz} \rightarrow -8i \cos \phi/N\pi r^2 \sin^2 \theta$, or, written in general Cartesian co-ordinates including the $y$-component,
\begin{equation}
F_{iz} \rightarrow \frac{-8i x_i}{N\pi r_\perp^3} 
\end{equation}
Where $r_\perp = \sqrt{x^2+y^2}$ is the perpendicular distance from the Wilson line. This agrees with (5.19) in \cite{Aharony:2022ntz}, if we take the charge $q=2$, when we are considering the bosonic theory instead of the fermionic, which could be seen from our earlier effective action for the gauge field which in this bosonic case differs from the fermionic by a factor of 2 (and recall that this paper writes the quantity with the indices reversed, $F_{zi}$, so that the negative appears).

For completeness, we also present the saddle-point value of $\lam(x)$. Recall from Section \ref{infspect} that in the case of an infinite Wilson line of charge $q$ we had $\lam \sim E^2 \sim \frac{q^2}{N^2}$. In the case of a Wilson line with endpoints, this behavior is quite different; the presence of the scalar field insertion implies that we obtain a nonzero $\lam$ already at $\sO(N^{-1})$. A simple way to see this is to consider defining it in terms of a 3-point function as in \eqref{Fdef3pt}:
\be
\langle \lam(x)\rangle_{W}= \frac{\langle \lam(x) \phi_{\beta}^{\dagger}(x_2)e^{i\int_{x_1}^{x_2} A}  \phi_{\al}(x_1) \rangle}{\langle \phi_{\beta}^{\dagger}(x_2)e^{i\int_{x_1}^{x_2} A} \phi_{\al}(x_1)\rangle}  = \frac{\langle \lam(x) \phi_{\beta}^{\dagger}(x_2)\phi_{\al}(x_1) \rangle}{\langle \phi_{\beta}^{\dagger}(x_2)\phi_{\al}(x_1)\rangle} + \sO(N^{-2})\label{Odef3pt}
\ee
Here we see that -- contrary to the case of the gauge field -- the leading order contribution is actually determined by conformal invariance in the undeformed theory, and a simple calculation using the usual $\mathbb{CP}^{N-1}$ Feynman rules allows one to compute the prefactor as:
\be
\langle \lam(x)\rangle_{W} = \frac{4}{N \pi^2 |x|^2} + \sO(N^{-2}) \label{lamprof} 
\ee
To leading order $\lam$ is essentially blind to the Wilson line, and is simply describing the physics of the self-interaction of the $\phi$ fields that are living on the endpoints of the line. A detailed description of the derivation of \eqref{lamprof} and a diagrammatic way to think about the relationship to the earlier computation can be found in Appendix \ref{app:lamsaddle}. 
\subsection{Spectrum of endpoint} 
With an eye towards studying the state-operator correspondence, we now quantize the theory on $S^2 \times \mathbb{R}$, where the $S^2$ in question surrounds one of the endpoints. We are thus looking at the spectrum of the theory with Hamiltonian
\be
H_{\mathrm{endpoint}} = H_{\mathrm{CFT}} + iA_{\tau}(\th=0) \label{hamendpoint}
\ee
i.e. where the Hamiltonian has only a single deformation at the North pole of the sphere.

To proceed, we first consider the ground state. The equation of motion of $A_{\tau}$ arising from varying \eqref{hamendpoint} tells us that the $U(1)$ gauge charge in the scalar sector must be $1$, and thus that the ground state which we denote $\ket{\phi}$ has a single scalar quantum in it. In the strict $N \to \infty$ limit, the energy of this state is obtained by setting $n=m=0$ in \eqref{freespec}, and the state has energy $\Delta_\phi= \ha$ and a spatially constant wavefunction $\Phi(\th,\phi) = \frac{1}{\sqrt{4\pi}}$. By the state-operator correspondence, this state simply maps to the operator $\Phi_{\al}$ at the end of the Wilson line in \eqref{repeatWilson}. 

We may now consider how to proceed to first order in $N^{-1}$. Here we can identify the expectation values of the gauge field and $\lambda$ by conformally mapping \eqref{Aans} and \eqref{lamprof} to $S^2 \times \mathbb{R}$. 
\begin{equation}
\bra{\phi} A^{S^2 \times \mathbb{R}}(x) \ket{\phi} = \frac{8}{N \pi}\left(1-\frac{\theta}{\pi}\right) \left(\cot\theta d\tau -d\theta\right)  \qquad \bra{\phi} \lambda^{S^2 \times \mathbb{R}} \ket{\phi} = \frac{4}{N \pi^2} \label{bgendpoint} 
\end{equation}
where both results are correct only to leading order in $\sO(N^{-1})$. 

We would also like to compute the shift in the dimension $\Delta_\phi$. This is more difficult; in principle it would be done through second-order quantum-mechanical perturbation theory, which tells us that the energy is shifted by
\be
\delta \Delta_\phi = \sum_{\Delta' \neq \Delta_\phi} \frac{|\bra{\Delta'} H_{int} \ket{\Delta_{\phi}}|^2}{\Delta_{\phi}-\Delta'} \label{2pt} 
\ee
where the sum runs over energy eigenstates of the infinite-$N$ Hamiltonian. To understand the structure of the interaction Hamiltonian $H_{int}$, consider writing $A = \frac{\hat{A}}{\sqrt{N}}$ and $\lam = \frac{\hat{\lam}}{\sqrt{N}}$ so that the kinetic term in \eqref{Fprof} (and its analogue for $\lam$) are canonically normalized. The interaction terms coupling $A$, $\lam$ to $\phi$ then take the form
\be
H_{int} \sim \frac{1}{\sqrt{N}} \int d\Om_2 \le(\hat{A} \phi_{\al} ^{\dagger} \p \phi_{\al} + \hat{\lam} \phi^{\dagger}_{\al} \phi_{\al}\ri)
\ee
and thus the square of the mixed matrix elements in \eqref{2pt} and the shift of $\delta \Delta_{\phi}$ are of order $N^{-1}$. We can see that the important matrix elements are those which relate a single-particle $\phi$ state to one containing both a $\phi$ and an $A$ or $\lam$ quanta. In principle one could perform the necessary perturbation theory calculations using the formalism for the $\mathbb{CP}^{N-1}$ model on the sphere worked out in e.g. \cite{Dyer:2015zha}; in practice it is much simpler to perform a computation on $\mathbb{R}^3$, which we do in the next section. In particular, one can imagine cutting open the Feynman diagrams in Figure \ref{fig:abcfeynmandiagrams}  to access intermediate states involving either a $\phi$ quanta or one including in addition a $A$ or $\lam$ quanta, i.e. the states expected to be important from above. 

% This approach transfers the difficulty of computing matrix elements in the $S^2 \times \mathbb{R}$ picture to computing Feynman diagrams involving the Wilson line. In this latter task we have prior results such as those in \cite{kaul2008quantum}, which already provide some of the quantities we need to combine to calculate the shift in the conformal dimension. This, along with the existence of straightforward diagrammatic techniques in this theory, makes the approach that follows in the next section a more straightforward calculation than that of matrix elements in this picture.

\section{Conformal dimension of endpoint} \label{sec:confdim} 

In this section we will finally calculate the conformal dimension $\Delta_{\phi}$ of the endpoint to $\sO(N^{-1})$. We do this by explicitly computing the expectation value of a Wilson line extending in a straight line along the $z$ axis from $z_2$ to $z_1$, i.e. as
\be
W(z_1,z_2)_{\al\beta} = \phi_{\beta}^{\dagger}(z_2)\exp\le(i\int_{z_1}^{z_2} A\ri) \phi_{\al}(z_1) \ . 
\ee
This configuration is actually equivalent -- under a conformal inversion about the point half-way between the two lines -- to the configuration of two semi-infinite Wilson lines which are connected through infinity. 

\begin{figure}
    \centering
    \includegraphics[width=0.3\linewidth]{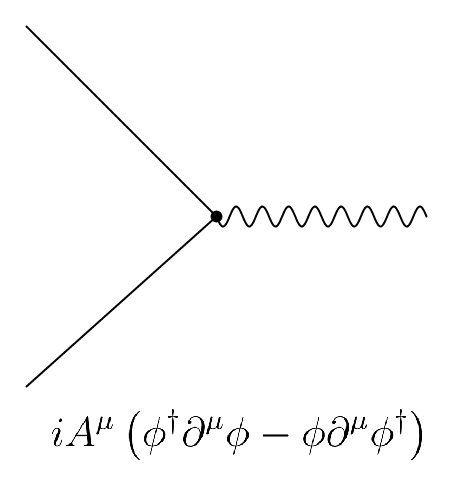}
    \caption{The scalar-gauge field interaction vertex, with Feynman rule.}
    \label{fig:interaction_vertex}
\end{figure}

\begin{figure}
\centering
\includegraphics[width=0.75\linewidth]{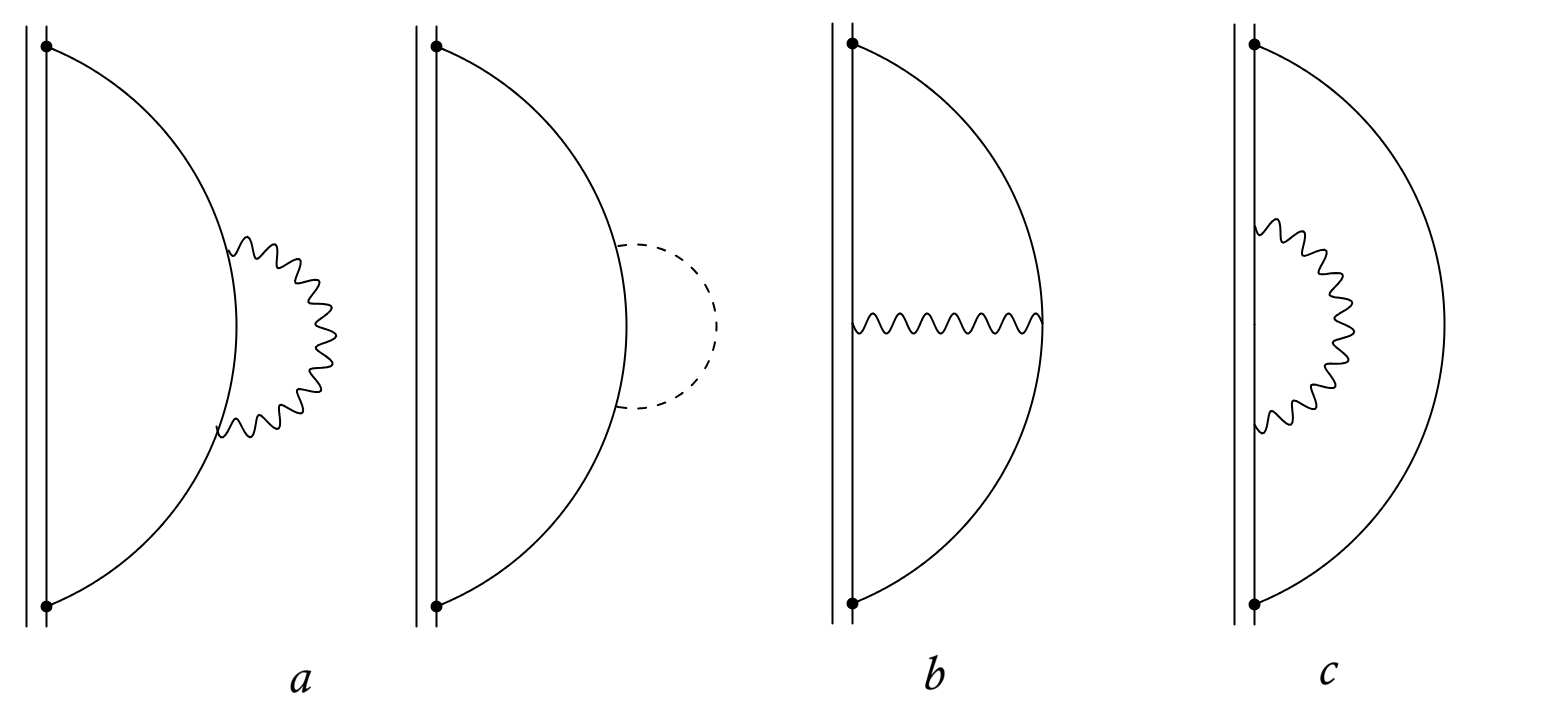}
\caption{Feynman diagrams showing the different contributions to the expectation value of the Wilson line with endpoints. Double line is the fixed external Wilson line, single line is propagator for $\phi_{\al}$, dashed line propagator for $\lam$, and wavy line propagator for $A$.}
\label{fig:abcfeynmandiagrams}
\end{figure}

We now explicitly compute the expectation value of this object as a function of $z_1 - z_2$ to $O(N^{-1})$. This dependence will determine the conformal dimension of the endpoint $\Delta_{\phi}$. We will require the gauge field propagator following from \eqref{eq:effective_double_puncture_action}; inverting the kinetic term using a gauge fixing parameter $\zeta$ as in \cite{kaul2008quantum, Benvenuti:2019ujm} we find
\begin{equation}\label{eq:effective_gauge_two_point_function}
\langle A_\mu (k) A_\nu (-k) \rangle = \frac{16}{N|k|} \left(\delta_{\mu \nu} - \zeta \frac{k_\mu k_\nu}{|k|^2}\right) 
\end{equation}
and Fourier transforming to position space we find:
\be
\langle A_\mu (x) A_\nu (0) \rangle  \equiv D_{\mu\nu}^{A}(x) = \frac{8}{\pi^2 N} \left(\frac{(1-\zeta)\delta_{\mu\nu}}{|x|^2}+2\zeta \frac{x_\mu x_\nu}{|x|^4}\right) \label{posspaceprop} 
\ee
Note that the value of $\zeta$ must drop out of all physical observables; this will be a useful check on our calculations. We will also require the scalar field propagator
\be
\vev{\phi^{\dag}(x)_{\al} \phi(0)_{\beta}} \equiv \delta_{\al\beta} D^{\phi}(x) = \delta_{\al\beta} \frac{1}{4\pi|x|}
\ee
The interaction vertex is illustrated in Figure \ref{fig:interaction_vertex}.\footnote{See \cite{Benvenuti:2019ujm} for a detailed discussion of the Feynman rules in variants of this theory.} 
We now compute the expectation value $\vev{W(z_1,z_2)_{\al\beta}}$. If we write 
\be
\Delta_{\phi} = \ha + \frac{\delta^{(1)}}{N} + \sO(N^{-2}) \label{del-expansion} 
\ee
as an expansion about the free dimension $\ha$, then we can expand the power law in powers of $N^{-1}$ to find: 
\be
\vev{W(z_1,z_2)_{\al\beta}} = \frac{\delta_{\al\beta}}{4\pi |z_1 - z_2|^{2\Delta_{\phi}}} \approx \frac{\delta_{\al\beta}}{4\pi|z_1 - z_2|}\le(1 - \frac{2 \delta^{(1)}}{N} \log |z_1 - z_2| + \cdots\ri)  \label{dim-expansion} 
\ee
and our task is thus to compute to first order in $N^{-1}$ the terms that have a logarithmic divergence in $z_1 - z_2$ and pick off the prefactor $\delta^{(1)}$.  It is now helpful to expand out the exponential as
\be
\vev{W(z_1,z_2)_{\al\beta}} = \vev{\phi_{\beta}^{\dagger}(z_2)\phi_{\al}(z_1)\le(1 + i \int dz A + \frac{i^2}{2} \int dz dz' A(z) A(z')+ \cdots \ri )}
\ee
To this order the evaluation results in the diagrams shown in Figure \ref{fig:abcfeynmandiagrams}. Note that each of them is individually gauge-variant, and none of them individually carries a meaningful physical interpretation. The fact that the gauge parameter $\zeta$ vanishes once they are added together is a useful check on the computation. This summed quantity is the meaningful, gauge-invariant result.

The two diagrams in part $(a)$ can be understood -- upon re-exponentiation -- as the expectation value of the two-point function of the bare $\phi$ field:
\be
\vev{\phi^{\dagger}_{\beta}(z_2) \phi_{\al}(z_1)} = \delta_{\al\beta}|z_1 - z_2|^{-2 \Delta_{\phi, \mathrm{bare}}}
\ee
This is a gauge-variant object, but its conformal dimension in a given gauge specified by $\zeta$ defined as in \eqref{eq:effective_gauge_two_point_function}  was worked out in \cite{kaul2008quantum} to be
\be
\Delta_{\phi, \mathrm{bare}} = \ha - \frac{2}{\pi^2 N} \le(3 + 2 \zeta\ri) \ . 
\ee
leading to a contribution to the anomalous shift defined in \eqref{del-expansion} of $\delta^{(1)}_a = -\frac{2}{\pi^2} (3 + 2\zeta)$. The computation captures how the self-energy of $\phi$ is renormalized by $\lam$ and $A$ lines. (To our knowledge the first computation of this critical exponent was in \cite{halperin1974first}, which worked in the gauge $\zeta = 1$). 

We now compute the diagram corresponding to part $(b)$ in Figure \ref{fig:abcfeynmandiagrams}. In position space the amplitude of interest takes the form
\be
I(z_1, z_2) = \int_{z_1}^{z_2} dw \vev{A_{z}(w) \phi^{\dagger}_{\beta}(z_2) \phi_{\al}(z_1)}
\ee
which we can express explicitly in terms of the appropriate propagators to find
\be
I(z_1,z_2) = i\int_{z_1}^{z_2} dw d^3 y D^{A}_{z\nu}(w,y)\le(D^{\phi}(z_1, y) \p^{\nu} D^{\phi}(y,z_2) - (z_1 \leftrightarrow z_2)\ri)
\ee
Due to its nonstandard kinematics this diagram is slightly nontrivial to evaluate. As usual for computing such anomalous dimensions it appears naively scale-invariant but contains UV divergences which contribute logarithmic terms of the form \eqref{dim-expansion}. It turns out to be easiest to perform the integral over the leg of the propagator attached to the Wilson line $w$ first and then look for UV divergences in the remaining integral, which appear when the bulk integration point $y$ coincides with either of the insertion points $z_1$ or $z_2$. The details are recorded in Appendix \ref{app:feynmancomp}; the upshot is that the contribution to the $N^{-1}$ correction to the anomalous dimension defined in \eqref{del-expansion} is: 
\be
\delta^{(1)}_{b} = -\frac{8}{\pi^2}(1-\zeta) \label{harddelta} 
\ee
Finally we compute the diagram corresponding to part $(c)$ in Figure \ref{fig:abcfeynmandiagrams}, corresponding to the disconnected product
\be
-\ha \vev{\phi^{\dagger}_{\beta}(z_2) \phi_{\al}(z_1)}\int dw_{a} dw_{b} \vev{A_{z}(w_a) A_{z}(w_b)}
\ee
This can be understood as the energetics of the interaction of the Wilson line with the field that it itself creates. This is relatively simple; we compute 
\be
\int_{z_1}^{z_2} dw_a \int_{z_1}^{z_2} dw_b D^{A}_{zz}(w_a, w_b) = \frac{8}{\pi^2 N}(1+\zeta) \int_{z_1}^{z_2} dw_{a} dw_{b} \frac{1}{|w_{a} - w_{b}|^2},
\ee
an integral whose universal part works out to be 
\be
-\frac{16}{\pi^2 N}(1+\zeta) \log\le(\frac{z_1 - z_2}{\ep}\ri)
\ee
Assembling the factors we find a contribution to the anomalous dimension which is 
\be
\delta^{(1)}_c = -\frac{4}{\pi^2} (1 + \zeta)
\ee
We can find the full shift by assembling $\delta^{(1)} = \delta^{(1)}_a + \delta^{(1)}_{b} + \delta^{(1)}_c = -\frac{18}{\pi^2}$. We thus find that the universal part of the two-point function takes the form
$\vev{W(z_1,z_2)_{\al\beta}} = \delta_{\al\beta} |z_1 - z_2|^{-2\Delta_{\phi}}$
where we have
\be
\boxed{\Delta_\phi = \frac{1}{2}-\frac{18}{\pi^2 N}+\mathcal{O}\left(\frac{1}{N^2}\right)}
\ee
This -- an explicit computation of the conformal dimension of the endpoint of the Wilson line -- is one of our main results. It provides a gauge-invariant definition of the dimension of a ``gauge-charged'' field (to order $1/N$); one must simply view this gauge-charged object as the endpoint of a line operator. 

The sign of the correction could perhaps have been anticipated: in the $S^2 \times \mathbb{R}$ picture one imagines sourcing the bulk electric field with a fixed external point charge at the north pole, which is then screened by a $\phi$ particle which is pulled forth from the vacuum. One would imagine that the energy of the $\phi$ quanta should be reduced from its free value to make it (energetically) worth its while to appear, and this is indeed what we find.

\section{Conclusion}

In this work we studied aspects of extended objects in QED$_3$ at its $\mathbb{CP}^{N-1}$ model fixed point for $N$ charged scalars. We began with an 
 infinite Wilson line, studied its stability properties in the $\mathbb{CP}^{N-1}$ model and discussed the state-operator correspondence for defect-localized operators. We then turned to a careful study of the endpoint of the Wilson line; here we presented a calculation of the conformal dimension of the endpoint to the first nontrivial order in $N^{-1}$, and also presented a formula for the field strength near the endpoint.   
 
 We briefly mentioned the existence of an OPE that could be used to sew together two open-ended Wilson lines. It would be very interesting to make further use of this technology, perhaps in the conformal bootstrap, a program initiated in \cite{Lanzetta:2024fmy, Lanzetta:2025xfw}. It would also be interesting if the critical exponent that we calculated above -- a new piece of conformal data -- could be verified through numerical lattice simulations. Indeed, though our calculations do not directly apply for $N = 1$, the theory there is related via particle-vortex duality to the XY model, and in that case our endpoint operator seems to be related to a vortex-ending monopole operator which could in principle be simulated on the lattice, perhaps using the numerical techniques of \cite{Zhou:2023fqu,Zhu:2022gjc,Voinea:2024ryq,Hu:2023ghk}.

Finally, an interesting further direction to explore would be the study of the gravitational dual of a defect with endpoint; in the CFT$_3$ case studied here this would correspond to a 2d string worldsheet hanging down into AdS$_4$ and ending on a massive particle worldline in the bulk. There are multiple levels of study possible; one could treat the string worldsheet and particle as fundamental objects in the bulk, or one could presumably regularize their internal structure in terms of an appropriate backreacted gravitational solution, as was done for a 1d boundary defect in \cite{Dias:2013bwa}.

\vspace{2cm} 
{\bf Acknowledgments:} NI and NN are supported in part by STFC grant number ST/T000708/1. This work was supported by a grant from the Simons Foundation (PD-Pivot Fellow-00004147, NI). NN was supported by a Delta ITP Visiting PhD Student Fellowship. 

\begin{appendix}

\section{Derivation of bosonic effective gauge field action}\label{app:deriv_effective_action}
Here we integrate out the scalar field $\phi$ from scalar QED to obtain the effective gauge action. We assume $A \sim 1/N$ as has been shown to be the case in our calculations throughout the paper by a systematic and consistent $1/N$ expansion; then we can write
\begin{equation}
e^{-S_{eff}[A]} = Z_\phi\left(1+ \frac{1}{2}\sum_{\alpha, \beta} \int d^3 x d^3 y \, \langle j^\mu_\alpha(x) j^\nu_\beta(y) \rangle A_\mu(x) A_\nu(y)\right) + \mathcal{O}\left(1/N^2\right)
\end{equation}h
Where$Z_\phi$ is the free scalar theory partition function, required to be multiplied above to reproduce the correct definition of the correlation function. Now by Wick contraction, and in agreement with the conformal two-point function for a spin-1 field,
\begin{flalign}
\sum_{\alpha, \beta} \langle j_\alpha^\mu(x) j_\beta^\nu(y) \rangle &= 2N_b \left(\frac{(x-y)^\mu (x-y)^\nu}{16\pi^2|x-y|^6} + \frac{\eta^{\mu\nu}}{16\pi^2|x-y|^4} - 3\frac{(x-y)^\mu (x-y)^\nu}{16\pi^2|x-y|^6}\right) \nonumber \\ &= \frac{N_b}{8\pi^2} \left( \frac{\eta^{\mu\nu}}{|x-y|^4}-2\frac{(x-y)^\mu (x-y)^\nu}{|x-y|^6}\right)
\end{flalign}
Where we've used the scalar free propagator
\begin{equation}
\langle \phi^\dag(x) \phi(y) \rangle = \frac{1}{4\pi |x-y|}
\end{equation}
Now we want to Fourier transform everything to momentum space. We get
\begin{equation}
e^{-S_{eff}[A]} = Z_\phi \left(1+\frac{N_b}{16\pi^2} \int \frac{d^3 k}{(2\pi)^3} \, \left( \frac{\pi^2 k^\mu k^\nu }{2|k|} - \pi^2 \frac{\eta^{\mu\nu}}{2}|k| \right) A_\mu(k) A_\nu(-k)\right) + ...
\end{equation}
Using the standard Fourier transform result
\begin{equation}\label{eq:skenderis_formula}
\int d^3 x \, \frac{e^{-ik \cdot x}}{|x|^{2n}} = \frac{\pi^{3/2} 2^{3-2n} \Gamma \left(\frac{3}{2} - n \right)}{\Gamma(n)} |k|^{2n - 3}
\end{equation}
from which we take the particular case
\begin{equation}
\int d^3 x \, \left(-2\frac{\eta^{\mu\nu}}{|x|^4}+\frac{8x^\mu x^\nu}{|x|^6}\right)e^{-ik \cdot x} = - \frac{2\pi^2 k^\mu k^\nu }{|k|}
\end{equation}
Therefore we have
\begin{flalign}
S_{eff}[A] &= - \log Z_\phi - \frac{N_b}{16\pi^2}\int \frac{d^3 k}{(2\pi)^3} \, \left( \frac{\pi^2 k^\mu k^\nu }{2|k|} - \pi^2 \frac{\eta^{\mu\nu}}{2}|k| \right) A_\mu(k) A_\nu(-k) + ... \nonumber \\ &= \text{const} + \frac{N_b}{32}\int \frac{d^3 k}{(2\pi)^3} \, \left( \eta^{\mu\nu} - \frac{ k^\mu k^\nu }{|k|^2} \right) |k| A_\mu(k) A_\nu(-k) + ...
\end{flalign}

\section{Using on-shell eigenvalues to calculate functional determinant}\label{app:onshell} 

In this section we briefly discuss a method for obtaining functional determinants using only on-shell data. For concreteness consider the scalar field wave equation described around \eqref{scalac} 
\be
\le(-D_{\mu} D^{\mu} + \lam + \frac{1}{4}\ri) \phi = 0
\ee
We are interested in computing a functional determinant of the operator $\Box \equiv (-D_{\mu} D^{\mu} + \lam + \frac{1}{4}$, and if we write
\be
\Box \phi = \Lam \phi
\ee
with $\Lam$ the eigenvalues of $\Box$, then this determinant is formally
\be
\mathcal{D} \equiv \log \det \Box = \Tr \log \Lam
\ee
Let us unpack this more carefully: we are working on $S^2 \times \mathbb{R}$, and thus the eigenvalues will depend on one continuous quantum number $\om$ (conjugate to translations along $\mathbb{R}$) and two discrete quantum numbers $m$ and $n$, where we take $m$ to be conjugate to translations along $\phi$ and where $n$ describes the discrete harmonics in the $\th$ direction. Thus we have
\be
\mathcal{D}(\lam) = \Tr \log \Lam = \sum_{m,n} \int \frac{d\om}{(2\pi)} \log \Lam(\om,m,n;\lam)
\ee
where we have made explicit the dependence on various quantities. 

It is convenient to compute
\be
\frac{d}{d\lam}\mathcal{D}(\lam) = \sum_{m,n} \int \frac{d\om}{(2\pi)} \frac{1}{\Lam} \frac{d\Lam}{d\lam}
\ee
The integral is taken over the real $\om$ axis. Now let us consider completing the contour above; in this case, the integral will receive contributions only from the zeros of $\Lam$ in the complex plane. These zeros are precisely on the frequencies $\om^*_{m,n}(\lam)$ when we have a solution to the on-shell wave equation $\Box \phi = 0$. In the neighbourhood of those solutions we have
\be
\Lam(\om,m,n;\lam) \sim \om - \om^*_{m,n}(\lam)
\ee
and thus near each singularity we can write 
\be
\frac{d}{d\lam}\mathcal{D}(\lam) \sim \sum_{m,n} \int \frac{d\om}{(2\pi)} \frac{1}{\om - \om^*_{m,n}(\lam) } \frac{d \om^*_{m,n}(\lam)}{d\lam}
\ee
Performing the integral by residues we thus find
\be
\frac{d}{d\lam} \mathcal{D}(\lam) = -i\sum_{m,n} \frac{d \om^*_{m,n}(\lam)}{d\lam}
\ee
Thus we see that the dependence of the determinant on any parameter can be understood from the derivative of the on-shell frequencies with respect to that parameter. This expression is used in \eqref{contint} in the bulk of the paper.

\section{Fourier transform details}
Here we describe some of the details in computing the equation of motion for the half-Wilson line in Fourier space. Since we are working in momentum space, we'll need the Fourier transform of this half-line source. Recalling the definition of the Heaviside step function
\begin{flalign}
\Theta(z) &= 0 \, | \, x<0 \nonumber \\ &= \frac{1}{2} \, | \, x=0 \nonumber \\ &= 1 \, | \, x>0
\end{flalign}
From this it is straightforward to see that $\Theta(z)+\Theta(-z)=1$ for all $z$. We use this fact in calculating the Fourier transform as follows:
\begin{flalign}\label{eq:first_part_heaviside}
\int d^3 x \, i\delta^{z\mu} \delta^{(2)} (x,y) \Theta(z) e^{-i\textbf{k} \cdot \textbf{x}} &= i\delta^{z\mu} \int_{-\infty}^\infty dz \, \Theta(z) e^{-ik_z z} \nonumber \\ &= i\delta^{z\mu} \left(\left[\frac{\Theta(z)}{-ik_z} e^{-ik_z z} \right]^\infty_{-\infty} - \frac{i}{k_z}\right)
\end{flalign}
Where on the second line, we've integrated by parts. Let's now use the fact above to also write this as
\begin{flalign}
&i\delta^{z\mu} \left(\left[\frac{1}{-ik_z} e^{-ik_z z} \right]^\infty_{-\infty} - \left[\frac{\Theta(-z)}{-ik_z} e^{-ik_z z} \right]^\infty_{-\infty} - \frac{i}{k_z}\right) \nonumber \\ &= i\delta^{z\mu} \left(2\pi \delta(k_z) +\left[\frac{\Theta(z)}{-ik_z} e^{ik_z z} \right]^{\infty}_{-\infty} - \frac{i}{k_z}\right)
\end{flalign}
Where to evaluate the first term in square brackets, we've used the definition of the Fourier transform of $1$ by convention:
\begin{equation}
\int dz \, e^{-ik_z z} = \left[\frac{1}{-ik_z} e^{-ik_z z} \right]^\infty_{-\infty} = 2\pi \delta(k_z)
\end{equation}
Now adding both expressions, we deduce
\begin{flalign}\label{eq:clear_heaviside_deriv}
2\int d^3 x \, i\delta^{z\mu} \delta^{(2)} (x,y) \Theta(z) e^{-i\textbf{k} \cdot \textbf{x}} &= i\delta^{z\mu} \left(2\pi \delta(k_z) +2\left[\frac{\Theta(z)}{-ik_z} \cos (k_z z) \right]^{\infty}_{-\infty} - \frac{2i}{k_z}\right) \nonumber \\ \Rightarrow \int d^3 x \, i\delta^{z\mu} \delta^{(2)} (x,y) \Theta(z) e^{-i\textbf{k} \cdot \textbf{x}} &=i\delta^{z\mu} \left(\pi \delta(k_z) +\lim_{z\rightarrow \infty}\left[\frac{1}{-ik_z} \cos (k_z z) \right] - \frac{i}{k_z}\right)
\end{flalign}
Now, the standard choice of principal value of the integral sets the term in square brackets to $0$ (such a choice must generally be made for Fourier transforms of distributions, like this). 

\section{Saddles of $\lambda$ in the presence of Wilson lines with endpoints}\label{app:lamsaddle} 
In this appendix, we present calculations of the saddle point of the Hubbard-Stratonovich field $\lambda$, in the presence of an infinite Wilson line, and in the presence of Wilson lines with endpoints. The methods are analogous to those already employed to find saddles of $F_{\mu\nu}$, and we use both path integrals and Feynman diagrams.

As in computing the saddles of $F_{\mu\nu}$, we can expand all terms that are $\mathcal{O}(1/N_b)$. For notational convenience, we write the entire path integral as $Z_\phi \left(1 + S_{int}[A,\lambda]\right) e^{-S[A]}e^{i\left(\int_{-\infty}^\infty dz \, A_z(z)\right)}$ (since the 1 from the exponential just gives the free scalar partition function $Z_\phi$), where $S_{int}[A,\lambda]$ is entirely $\mathcal{O}(1/N_b)$. We can write
\begin{flalign}
S_{eff}[A,\lambda] &= S[A] - \left(i\int_{-\infty}^\infty dz \, A_z(z)\right) - \frac{N_b}{\frak{g}} \int d^3 x \, \lambda(x) + \text{const.} + \log \left(1 + S_{int}[A,\lambda]\right) \nonumber \\
&= S[A] - \left(i\int_{-\infty}^\infty dz \, A_z(z)\right)- \frac{N_b}{\frak{g}} \int d^3 x \, \lambda(x) + \text{const.} + S_{int}[A,\lambda] - \frac{1}{2} S_{int}[A,\lambda]^2 + ...
\end{flalign}
We can expand and write down the leading-order nonvanishing terms:
\begin{align}
S_{int}[A,\lambda] = \sum_{\alpha} \int d^3 x \, \lambda(x) \langle \phi^\dag_\alpha(x) \phi_\alpha(x) \rangle + ... \nonumber \\ ... + \sum_{\alpha, \beta} \left(\frac{1}{2} \int d^3 x \int d^3 x' \, \langle j_\alpha^\mu(x) j_\beta^\nu (x') \rangle A_\mu(x) A_\nu(x') + \text{other terms in $A$}\right) + ...\nonumber \\ ...+ \sum_{\alpha, \beta} \frac{1}{2} \int d^3 x \int d^3 x' \, \lambda(x) \lambda (x') \langle \phi_\alpha(x) \phi^\dag_\alpha(x) \phi_\beta(x') \phi^\dag_\beta(x') \rangle + ... \nonumber \\ ... + \sum_{\alpha, \beta, \gamma} \frac{1}{2} \int d^3 x \int d^3 y \int d^3 z \, \lambda  (z) \langle \phi_\gamma^\dag(z) \phi_\gamma(z) j_\alpha^\mu(x) j_\beta^\nu (y) \rangle A_\mu(x) A_\nu(y) + \mathcal{O} \left(1/N_b^4\right)
\end{align}
Thus, differentiating $S_{eff}[A,\lambda]$ with respect to $\lambda$ gives the saddle-point equation for $\lambda$ (where we use the saddle point expression for $A$):
\begin{align}\label{eq:fullsaddleforlambda}
 \sum_{\alpha} \left(\langle \phi^\dag_\alpha(x) \phi_\alpha(x) \rangle - \frac{N_b}{\frak{g}} \right) + \sum_{\alpha, \beta} \int d^3 x' \, \lambda(x') \langle \phi_\alpha(x) \phi^\dag_\alpha(x) \phi_\beta(x') \phi^\dag_\beta(x') \rangle + ... \nonumber \\ ... + \sum_{\alpha, \beta, \gamma} \frac{1}{2} \int d^3 y \int d^3 z \,  \langle \phi_\gamma^\dag(x) \phi_\gamma(x) j_\alpha^\mu(y) j_\beta^\nu (z) \rangle A_\mu(y) A_\nu(z) + \mathcal{O}\left(1/N_b^2\right) 
\end{align}
Note that when we tune this theory to criticality, we impose that $\sum_\alpha \langle\phi_\alpha^\dag \phi_\alpha\rangle=N_b/\frak{g}_c$ at the critical point. Hence the first term disappears.

Note that at each order in $1/N_b$, the $\log$ will ensure that disconnected contributions from this expectation value of $\phi^\dag_\alpha(x) \phi_\alpha(x)$ in larger correlators will always cancel, meaning that we still need only consider the usual connected, free scalar correlators as before.

Thus the saddle-point equation for $\lambda$ to leading order in $1/N_b$ is given by:
\begin{flalign}
\sum_{\alpha, \beta} \int d^3 x' \, \lambda(x') \langle \phi_\alpha(x) \phi^\dag_\alpha(x) \phi_\beta(x') \phi^\dag_\beta(x') \rangle \nonumber \\ = - \frac{1}{2} \sum_{\alpha, \beta, \gamma} \int d^3 y \int d^3 z \,  \langle \phi_\gamma^\dag(x) \phi_\gamma(x) j_\alpha^\mu(y) j_\beta^\nu (z) \rangle A_\mu(y) A_\nu(z).
\end{flalign}
The equation above may be written in integral form, but is difficult to solve directly for $\lambda$. However, from the $SO(2)$ symmetry around the defect implicit in the saddle point profile of $A_\mu$ that we must substitute into this equation, and from the required weight of $\lambda$ being 2, it is clear that it must take the form $\lambda \sim \frac{E^2}{\sin^2\theta}$ in the $S^2 \times \mathbb{R}$ picture, as previously derived.

We can work identically in the single puncture case. However, here we must be more careful: there is no reason to expect $\lambda$ to be $\mathcal{O}(1/N_b^2)$, and indeed we will see that it is not; it will instead turn out to be $\mathcal{O}(1/N_b)$, a fact that will illuminate the utility of the OPE. 

Therefore let us write:
\begin{equation}
\lambda = \frac{\lambda^{(-1)}}{N_b} + \mathcal{O} \left(1/N_b^2\right)
\end{equation}
This can be checked for consistency at the end. Then, carrying out the same computation of the effective action with the insertion of Wilson lines ending in points, we have:
\begin{flalign}
e^{-S_{eff}[A,\lambda]} &= e^{-S[A]} e^{i\int_{0}^\infty dz A_z(z)}  \int \prod_{i=1}^{N_b} \lim_{z \rightarrow \infty}\mathcal{D}\phi_i \, \frac{\phi_{\bar{\rho}}(0) \phi_{\bar{\rho}}^\dag(z)}{\langle \phi_{\bar{\rho}}(0) \phi_{\bar{\rho}}^\dag(z) \rangle}e^{-\sum_{\alpha = 1}^{N_b} S[\phi_\alpha]} \times \nonumber \\ &\times e^{-\int d^3 x \,  \left( \left(\sum_{\alpha = 1}^{N_b} j^\mu_\alpha A_\mu + \lambda \phi^\dag_\alpha \phi_\alpha \right) - \frac{\lambda}{\frak{g}}  N_b \right)}
\end{flalign}
Here, the barred subscript $\bar{\rho}$ indicates that we are taking a single scalar field at each end of the Wilson line, both of the same flavour, with \textit{no sum} over $\bar{\rho}$ (this is exactly what we did in the computation of the saddle of $F_{\mu\nu}$. 

As before, the leading order term in the saddle-point equation for $\lambda$ gives us:
\begin{equation}
\sum_\alpha \langle \phi^\dag_\alpha(x) \phi_\alpha(x) \rangle - \frac{N_b}{\frak{g}} = 0.
\end{equation}

At the next order, things are different from the double puncture case. A new term arises from Wick contracting the $\phi_\alpha^\dag \phi_\alpha$ with the endpoints; that is, the connected part:
\begin{equation}
\lim_{z \rightarrow \infty} \sum_\alpha \frac{\langle \phi_{\bar{\rho}}(0) \phi_\alpha^\dag(x) \phi_\alpha(x) \phi_{\bar{\rho}}^\dag(z) \rangle_c }{\langle \phi_{\bar{\rho}}(0) \phi_{\bar{\rho}}^\dag(z) \rangle} = \frac{1}{4\pi|x|}.
\end{equation}
Thus, the saddle-point equation at next-to-leading order is:
\begin{align}
\frac{1}{4\pi|x|} = \sum_{\alpha, \beta} \int d^3 x' \, \lambda^{(-1)}(x') \langle \phi_\alpha(x) \phi^\dag_\alpha(x) \phi_\beta(x') \phi^\dag_\beta(x') \rangle_{c}.
\end{align}
An interesting observation from this single-puncture saddle point equation is that the saddle of $\lambda$ is essentially 'blind' to the saddle value of $A$: it is as though the Wilson line were not present, or put more accurately, its effects compared to those of the bare scalars are suppressed by a power of $1/N_b$. 

In fact, this integral equation may be solved using the same Fourier integrals as earlier employed in the calculation of the saddle of $F_{\mu\nu}$ \eqref{eq:skenderis_formula}, and we obtain as our answer for the saddle-point of $\lambda$ in the single-puncture case:
\begin{equation}\label{eq:singlepuncturelambdafull}
\tilde{\lambda}(k) = \frac{8}{N_b |k|} + \mathcal{O} \left(1/N_b^2\right) \Rightarrow \lambda(x) = \frac{4}{N_b \pi^2 |x|^2} + \mathcal{O} \left(1/N_b^2\right)
\end{equation}
We can also transform this result to the $S^2 \times \mathbb{R}$ picture, recalling that $\lambda$ has weight 2:
\begin{equation}
\lambda_{S^2 \times \mathbb{R}} = \frac{4}{N_b \pi^2}
\end{equation}

Consider also that this result agrees with the normalised expectation that we get from using conformal three-point functions:
\begin{equation}
\langle \lambda(x) \rangle \sim \lim_{z \rightarrow \infty}\frac{\langle \lambda(x) \phi^\dag(0) \phi(z) \rangle}{\langle \phi^\dag(0) \phi(z) \rangle} = \lim_{z \rightarrow \infty}\frac{\frac{1}{|z|^{-1}|x-z|^{2}|x|^{2}}}{\frac{1}{|z|}} = \frac{1}{|x|^{2}}.
\end{equation}
Which shows that the derived saddle has the correct positional argument.

\begin{figure}
    \centering
\begin{tikzpicture}

% First setup with a gap in the line
% Vertical line from (0, 2) to (0, 1)
\draw[thick] (0,2) -- (0,0.4);
\node at (-0.55, 0.4) {$\phi^\dag(x)$};
% Vertical line from (0, -1) to (0, -2)
\draw[thick] (0,-0.4) -- (0,-2);
\node at (-0.55, -0.4) {$\phi(y)$};
% Black dots at (0,1) and (0,-1)
\filldraw (0,0.4) circle (2pt);
\filldraw (0,-0.4) circle (2pt);
% Black dot beside the line, shifted right by 0.5 units
\filldraw (-1.2, -0.65) circle (2pt);
\node at (-1.5, -0.35) {$\lambda(z)$};

% Second setup with an unbroken line
% Shifted right by 2 units for separation
\begin{scope}[xshift=4cm]
    \fill[gray!10] (-2.5, -2.5) rectangle (0.8, 2.5);
    % Continuous vertical line from (0, 2) to (0, -2)
    \draw[thick] (0,2) -- (0,-2);
    % Black dot beside the line, shifted right by 0.5 units
    \filldraw (-1.2, -0.65) circle (2pt);
    \node at (-1.4, -0.35) {$\lambda$};
\end{scope}
\begin{scope}[xshift=8cm]
    \fill[gray!20] (-2.5, -2.5) rectangle (0.8, 2.5);
    % Continuous vertical line from (0, 2) to (0, -2)
    \draw[thick] (0,2) -- (0,-2);
    % Black dot beside the line, shifted right by 0.5 units
    \filldraw (0, 0) circle (2pt);
    \node at (-0.55, 0) {$\phi^\dag \phi$};
    \filldraw (-1.2, -0.65) circle (2pt);
    \node at (-1.4, -0.35) {$\lambda$};
\end{scope}

\end{tikzpicture}
    \caption{The OPE applied to a Hubbard-Stratonovich field $\lambda$ in the vicinity of coaxial separated Wilson lines ending in points. We show the first two contributions in the OPE; further contributions will come from higher weight primaries localised on the defect.}
    \label{fig:application_of_ope}
\end{figure}
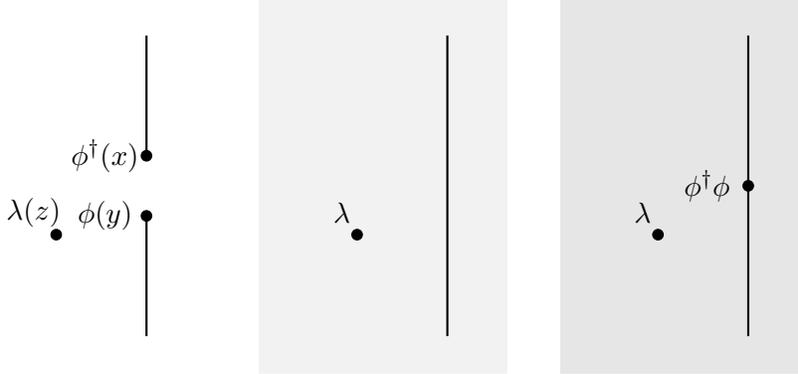

One could also imagine more generally a setup where a bulk insertion of $\lambda$ is placed next to two coaxial Wilson lines ending in points separated by a distance; this is illustrated in Figure \ref{fig:application_of_ope}, and provides another simple application of our OPE. Here, we present a short calculation to exhibit the conformality of the expectation value of this setup.
Considering the setup in Figure \ref{fig:application_of_ope}, then, we first find the positional form of the expectation value of this setup using Feynman rules in momentum space, and then show that this result is conformal.

\begin{figure}
    \centering
\begin{tikzpicture}
    % Define the vertex
    \node[circle,fill=black,inner sep=1.5pt] (v) at (0,0) {};

    % Draw the solid line at 45 degrees upwards to the right
    \draw[-] (v) -- (-2,-2) node[midway, above right] {};

    % Draw the solid line at 45 degrees downwards to the left
    \draw[-] (v) -- (-2,2) node[midway, above left] {};

    % Draw the dashed line horizontally to the right
    \draw[dashed,->] (v) -- (3,0) node[midway, below] {};
\end{tikzpicture}
    \caption{The 3-point function $ \langle \lambda \phi^\dag \phi \rangle$ in momentum-space.}
    \label{fig:lambda_momentum_space}
\end{figure}
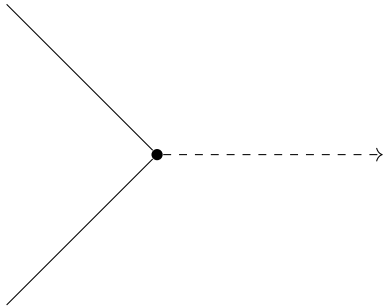

The leading order Feynman diagram in momentum space is the one depicted in Figure
\ref{fig:lambda_momentum_space}. 

This is straightforward to evaluate, and gives the answer
\begin{equation}
\frac{-8\delta^{(3)} \left(p_1 + p_2 + p_3\right)}{N p_1^2 p_2^2 p_3^{-1}}
\end{equation}
We now wish to compare this to the conformal three-point function in momentum space presented in \cite{Bzowski:2013sza}. From (3.4) in that paper, we know that the conformal 3-point function for our case $\langle \lambda \phi^\dag \phi \rangle$ takes the following form as an integral in momentum space:
\begin{equation}\label{eq:three_point_as_triplek}
\frac{(2\pi)^3 c_{123} \pi^3 2^{11/2} \delta^{(3)}\left(p_1+p_2+p_3\right)}{\Gamma(0)\Gamma(\frac{3}{2})\Gamma(\frac{3}{2})\Gamma(-\frac{1}{2})} \times p_1^{-1} p_2^{-1} p_3^{\frac{1}{2}} \int_0^\infty dx \, x^{\frac{1}{2}} K_{-1}(p_1 x) K_{-1}(p_2 x) K_{\frac{1}{2}}(p_3 x)
\end{equation}
For $c_{123}$ the conformal OPE data. This is a generic expression, so we are not concerned with the specific value of this coefficient, but rather the positional form of the answer. Here, we need to treat the divergence in $\Gamma(0)$ carefully against the divergence in the integral; we do this by dimensional regularisation, following the methods of \cite{Bzowski:2013sza}.

Using the Mathematica package illustrated in \cite{Bzowski:2020lip}, it is possible to compute the regulated expression
\begin{equation}
\frac{(2\pi)^3 c_{123} \pi^3 2^{11/2} \delta^{(3)}\left(p_1+p_2+p_3\right)}{\Gamma(-\epsilon)\Gamma(\frac{3}{2})\Gamma(\frac{3}{2})\Gamma(-\frac{1}{2})} \times p_1^{-1} p_2^{-1} p_3^{\frac{1}{2}} \int_0^\infty dx \, x^{\frac{1}{2}+\epsilon} K_{-1}(p_1 x) K_{-1}(p_2 x) K_{\frac{1}{2}}(p_3 x)
\end{equation}
The triple-K expression here (that is, everything to the right of the $\times$ sign) evaluated using \cite{Bzowski:2020lip} results in
\begin{equation}
-\frac{\sqrt{\frac{\pi}{2}}}{p_1^2 p_2^2 p_3^{-1} \epsilon} + \mathcal{O}(1)
\end{equation}
But now, noting that 
\begin{equation}
\frac{1}{\Gamma(-\epsilon)} = - \epsilon + \mathcal{O}(\epsilon^2)
\end{equation}
We see that overall the result for \eqref{eq:three_point_as_triplek} is of the correct form as calculated directly from the Feynman diagram:
\begin{equation}
\langle \phi^\dag(p_1) \phi(p_2) \lambda(p_3) \rangle = 512 c_{123} \pi^5 \frac{\delta^{(3)} \left(p_1 + p_2 + p_3\right)}{p_1^2 p_2^2 p_3^{-1}} + \mathcal{O}(\epsilon)
\end{equation}
as required. This dimensional regularisation in momentum space would have corresponded to a complex regularisation of divergences in the position space picture by subtracting Feynman diagrams. 

\section{Computation of Feynman diagram for conformal dimension} \label{app:feynmancomp} 

Here we present the details in the computation of the non-trivial Feynman amplitude for calculating the conformal dimension $\Delta_{\phi}$. The amplitude in question is
\be
I(z_1, z_2) = \int_{z_1}^{z_2} dw \vev{A_{z}(w) \phi^{\dagger}(z_1) \phi(z_2)}
\ee
The photon propagator \eqref{posspaceprop} has two parts which we study separately. For the first part -- i.e. arising from the $(1-\zeta)\delta_{\mu\nu}$ part of the photon propagator, we find 
\be
I(z_1,z_2) = \alpha i \int_{z_1}^{z_2} dw_{z} d^3  y\frac{i}{(y_z - w_z)^2 + |\vec{y}|^2} \le(\frac{z_1 - y_{z}}{((z_1 - y_z)^2 + |\vec{y}|^2)^{\frac{3}{2}}((z_2 - y_{z})^2 + |\vec{y}|^2)^{\frac{1}{2}}} - 1 \leftrightarrow 2\ri)
\ee
where $\al = \frac{8}{\pi^2 N} \frac{1}{(4\pi)^2} (1-\zeta)$ 
Here we have split the bulk integration coordinate into directions parallel and perpendicular to the Wilson line respectively, i.e. $y = (y_{z}, \vec{y})$. For notational simplicity we set $z_1 = 0$ and $z_2 = L$; here $L$ is the distance between the two endpoints of the Wilson line. We then perform the integral over $w_{z}$ to find
\be
I = -2\pi \al i \int dy_{z} d|\vec{y}| |\vec{y}| \frac{L \left(y_{z} (L-y_{z})+|\vec{y}| ^2\right) \left(\tan ^{-1}\left(\frac{L-y_{z}}{|\vec{y}| }\right)+\tan ^{-1}\left(\frac{y_{z}}{|\vec{y}| }\right)\right)}{|\vec{y}|  \left(|\vec{y}| ^2+y_{z}^2\right)^{3/2} \left((L-y_{z})^2+|\vec{y}| ^2\right)^{3/2}}
\ee
The resulting integral is too complicated to be done in closed form; we would like to isolate the logarithmic divergences that contribute to the anomalous dimension. We expect these to arise when the bulk integration point $y$ coincides with one of the insertion points at the origin and at $(L,0,0)$. To that end let us move to spherical polar coordinates about the origin by $y_{z} = r \cos \th$, $|\vec{y}| = r \sin \th$. We then find
\be
I = 2\pi \al i \int_{\ep}^{\infty} r^2 dr d(\cos\th) \frac{L \csc (\theta ) (r \cos (2 \theta )-L \cos (\theta )) \left(\tan ^{-1}(\cot (\theta ))+\tan ^{-1}\left(\frac{\csc (\theta ) (L-r \cos (\theta ))}{r}\right)\right)}{r^{3}  \left(L^2-2 L r \cos (\theta )+r^2\right)^{3/2}}
\ee
Here we have introduces a UV cutoff $\ep$; it is clear that there is a log divergence arising from $r \approx \ep$. To extract the dependence it is convenient to rescale $r = L u$, in which case we find
\be
I = \frac{2 \pi \al i}{L} \int_{\frac{\ep}{L}}^{\infty} du d(\cos\th) \frac{\csc (\theta ) (u \cos (2 \theta )-\cos (\theta )) \left(\tan ^{-1}(\cot (\theta ))-\tan ^{-1}\left(\cot (\theta )-\frac{\csc (\theta )}{u}\right)\right)}{ u\left(u^2-2 u \cos (\theta )+1\right)^{3/2}}
\ee
The integrand is manifestly dimensionless and thus if it converged it would not contribute any $L$ dependence. Thus the full $L$ dependence will arise only from its failure to converge, i.e. from the logarithmic divergence at the origin, and a corresponding contribution from the {\it other} endpoint (which in these coordinates is at $u = 1, \th = 0$). Extracting the coefficient of the log divergence, performing the $\th$ integral, and and multiplying by two for the two endpoints we find
\be
I = -\frac{8\pi \al i }{L} \log\le(\frac{L}{\ep}\ri)
\ee
The remaining part of the integral -- arising from the $\frac{x_{\mu} x_{\nu}}{|x|^2}$ part of the propagator in \eqref{posspaceprop} -- can be done in the same way. As it turns out it does not contribute. We record some of the intermediate steps below. Here the $\mu$ index is tied to the external Wilson line and thus points always in the $z$ direction, and so the integral can be broken into a $zz$ block and a $zi$ block. The $zz$ block results in the following integral: 
\be
I^{zz} = -\tilde{\al}\int dw d^3y \frac{L (y_{z}-w_{z})^2 \left(|\vec{y}| ^2-y_{z} (y_{z}-L)\right)}{\left(|\vec{y}| ^2+y_{z}^2\right)^{3/2} \left((y_{z}-L)^2+|\vec{y}| ^2\right)^{3/2} \left(|\vec{y}| ^2+(y_{z}-w_{z})^2\right)^2}
\ee
and the $zi$ block results in:
\be
I^{zi} = -\tilde{\al}\int dw d^3y \frac{L |\vec{y}| ^2 (L-2 y_{z}) (y_{z}-w_{z})}{\left(|\vec{y}| ^2+y_{z}^2\right)^{3/2} \left((y_{z}-L)^2+|\vec{y}|^2\right)^{3/2} \left(|\vec{y}| ^2+(y_{z}-w_{z})^2\right)^2}
\ee
where $\tilde{\al} = \frac{8}{N\pi^2} \frac{1}{(4\pi)^2} \cdot 2\zeta$. Following the same steps as above, we find the these two contributions have identical logarithmic divergences with opposite sign, which thus cancel. Thus the divergence above is the full contribution to the integral from this diagram, corresponding to an anomalous shift of 
\be
\delta^{(1)}_{b} = -\frac{8}{\pi^2}(1-\zeta) \ . 
\ee
as reported in \eqref{harddelta}.

\end{appendix}

\bibliographystyle{utphys}

\bibliography{draft}

\end{document}